\documentstyle[preprint,eqsecnum,aps,epsf]{revtex}
\newif\iftightenlines\tightenlinesfalse
\tightenlines\tightenlinestrue

\def\eslt{E\llap/_T}
\def\to{\rightarrow}
\def\Re{{\cal R \mskip-4mu \lower.1ex \hbox{\it e}}\,}
\def\Im{{\cal I \mskip-5mu \lower.1ex \hbox{\it m}}\,}

\def\te{\tilde e}

\def\tb{\tilde b}
\def\tst{\tilde t}
\def\tt{\tilde t}
\def\ttau{\tilde \tau}
\def\tmu{\tilde \mu}
\def\tg{\tilde g}

\def\tnu{\tilde\nu}
\def\tell{\tilde\ell}
\def\tq{\tilde q}
\def\tw{\widetilde W}
\def\tz{\widetilde Z}
\def\cmsec{{\rm cm^{-2}s^{-1}}}
\def\fb{{\rm fb}}
\def\sgn{\mathop{\rm sgn}}
\def\mhf{m_{1/2}}
\def\jet{{\rm jet}}
\def\jets{{\rm jets}}
\def\dilepton{{\rm dilepton}}

\begin{document}
%
\preprint{\vbox{\baselineskip=14pt%
\rightline{FSU-HEP-951215}\break
   \rightline{UCD-95-45}
   \rightline{UH-511-837-95}
}}
\title{SIGNALS FOR MINIMAL SUPERGRAVITY\\
AT THE CERN LARGE HADRON COLLIDER II:\\
MULTILEPTON CHANNELS}
\author{Howard Baer$^1$, Chih-hao Chen$^2$, Frank Paige$^3$
 and Xerxes Tata$^{4,1}$}
\address{
$^1$Department of Physics,
Florida State University,
Tallahassee, FL 32306 USA
}
\address{
$^2$Dep't of Physics,
University of California,
Davis, CA USA
}
\address{
$^3$Brookhaven National Laboratory,
Upton, NY 11973 USA
}
\address{
$^4$Department of Physics and Astronomy,
University of Hawaii,
Honolulu, HI 96822 USA
}
\date{\today}
\maketitle
\begin{abstract}

We use ISAJET to perform a detailed study of the multilepton signals
expected from cascade decays of supersymmetric particle produced at
the CERN LHC.  Our analysis is performed within the framework of the
minimal supergravity model with gauge coupling unification and
radiative electroweak symmetry breaking.  We delineate the regions of
parameter space where jets plus missing energy plus 1, 2 (opposite
sign and same-sign dileptons), and 3 isolated lepton events should be
visible above standard model backgrounds. We find that if any $\eslt$
signal at the LHC is to be attributed to gluino and/or squark
production, and if $m_{\tg} \alt 1$~TeV, then several of these signals
must be simultaneously observable.  Furthermore, assuming
$10\,\fb^{-1}$ of integrated luminosity, we find that the reach for
supersymmetry in the 1$\ell + \jets +\eslt$ channel extends to
$m_{\tg}\sim 2300\ (1600)$ GeV for $m_{\tq}\sim m_{\tg}$ ($m_{\tq}
\sim 1.5m_{\tg}$), and exceeds the corresponding reach in the
$0\ell+\eslt$ channel.  We show that measurements of the various
topological cross sections, jet and $B$-hadron multiplicities in these
events, together with the charge asymmetry for single lepton and
same-sign dilepton events, and flavor asymmetry for opposite sign
dilepton events, serve to narrow the allowed range of underlying SUGRA
parameter values. We also delineate parameter regions where
signals with clean isolated dilepton (from slepton production) and
trilepton events (from chargino/neutralino production) are visible at
the LHC, and examine the extent to which these signals can be
separated from other SUSY sources.

\end{abstract}

\medskip



\section{Introduction and Motivation}

The search for supersymmetric particles is now an integral part
of all current, as well as future, experimental programs at
high energy colliders.
Aside from the many
attractive features of supersymmetry (SUSY), the impetus for these searches
comes from the fact that weak scale SUSY\cite{REV}, which is
introduced to ameliorate the fine tuning problem of the Standard
Model (SM), requires that the supersymmetric partners of SM
particles {\it must} be accessible to experiments that probe
the TeV energy scale. Thus, while experiments at LEP2 and
at the Tevatron (or its upgrades) may well discover sparticles,
a definitive search for supersymmetry can only be
performed\cite{DPF} at supercolliders
such as the CERN Large Hadron Collider (LHC) or, at electron-positron
linear colliders with $\sqrt{s}=500-1500$~GeV. There is general
agreement, based on detailed studies of SUSY signals
both within the more general Minimal Supersymmetric
Model (MSSM) framework\cite{EARLY,SDC,GEM,ATLAS}
as well as within the very attractive and economic supergravity\cite{FN1}
(SUGRA) GUT framework\cite{JLC,MUR,PHD,BCPT}, that weak scale SUSY will not
evade detection at these facilities\cite{FN2}.

The natural question then is: if we do see signals for new physics, can we
unravel their origin, and trace them to the production of supersymmetric
particles? At electron-positron colliders where the cleanliness of the
interaction environment allows for\cite{CHEN,JLC,MUR,FENG} the
precision measurement of at
least some of the properties of these particles
(mass, spin, decay patterns,...), this may be a straightforward exercise,
especially if the machine energy is increased incrementally, so that it is
possible to focus on just one new signal at a time.
At the LHC, however, the situation is much more complicated,
not only because of the messier environment, but
also because {\it all} new particles which are kinematically accessible will
simultaneously contribute to the signal: we will thus have the additional task
of sorting the supersymmetric signals from one another in order to
discover the nature of the new physics.

Some progress has already been made on the issue of identifying
the sparticle production processes that give rise to SUSY signals at the
LHC. For instance, it
has been shown\cite{TRILHC} that, with suitable cuts,
the clean $3\ell+\eslt$ signal from the production of charginos
and neutralinos via the reaction
$pp \to \tw_1\tz_2 \to \ell\nu\tz_1 + \ell'\bar{\ell}'\tz_1$
can not only be separated from SM backgrounds,
but also, that it can be isolated from other SUSY sources.
An observation of a signal in this channel would, therefore,
unambiguously point to $\tw_1\tz_2$ production as its source, at
least within the SUSY framework. It is, however, not always
possible to devise cuts to isolate a single source of SUSY events.
A detailed study of the signal characteristics may then help
to identify the sparticles producing the signal.
In a previous study\cite{BCPT}, hereafter
referred to as Paper I, we examined the reach of
the LHC in the multijet plus $\eslt$ channel and studied what
information could be obtained by a detailed study of this sample.
Assuming as usual that squarks cannot be much lighter than
gluinos, we
showed that if gluinos are lighter than about 750~GeV,
their mass could be extracted to 15 -- 25\%
by reconstructing multijet masses in opposite detector hemispheres.
Furthermore,
by measuring the mean jet multiplicity, $\langle n_{j} \rangle$,
which is observably larger if squarks are much heavier than gluinos,
it should be possible to distinguish the $m_{\tq} \simeq m_{\tg}$ case
from the one where squarks are substantially heavier than gluinos.
While the gluino mass would determine $\mhf$, $\langle n_{j} \rangle$
will at least enable us to decide whether $m_0$ is small or
comparable to $\mhf$, or much larger.

Are there other ways by which we can tell what is being produced
at the LHC? Also, is it possible
to test whether the minimal SUGRA framework adopted in Paper I
(as well as in many recent phenomenological
analyses\cite{JLC,MUR,BDKNT,LOPEZ,KAMON,BCMPT,TEVSTAR,MRENNA,LEP2} of SUSY)
can consistently account for all the observed signals, or
whether some of the underlying assumptions about the symmetries of
physics at the ultra-high scale need to be modified?
It has already been shown\cite{JLC,MUR} that the precision measurements that
are possible in the clean environment of $e^+e^-$
collisions will allow experimentalists to perform
incisive tests of the SUGRA framework at future Linear Colliders.
While it is not possible to perform similar measurements at
hadron colliders, the big advantage of the LHC over 500~GeV Linear
Colliders is that many more sparticle production processes
should be kinematically accessible, resulting in a large number
of potential observables. Since the minimal SUGRA GUT model
with radiative breaking of electroweak symmetry is completely
fixed by just four SUSY parameters: $m_0$ and $\mhf$, the universal scalar
mass and gaugino masses at the high scale $M_X \sim M_{GUT}$, the
SUSY breaking universal trilinear coupling $A_0$, and the parameter
$\tan\beta$ along with $\sgn\mu$, the consistency of the framework
can be tested by verifying that the rates and distributions
in all the observed
channels can be accommodated by a single choice of model parameters.
Even more ambitiously, one could ask whether it would be
possible to determine the underlying parameters from the observed
signals, and
we report the results of our preliminary attempt to do so in this paper.

We stress here that we do not mean to imply that the SUGRA framework
is the uniquely correct one. Indeed the sensitivity to the details of its
predictions should be examined, particularly when studying the
reach of future facilities. Nonetheless, it is an economic, attractive
and predictive framework, and it can be used as a guide for sparticle
masses and mixing patterns. Such a framework is needed since without
assuming anything other than the weak scale symmetries,
there are far too many parameters, making phenomenological analyses
intractable.

It should be clear from the preceding discussion that
a study of all possible signals as a function of SUGRA
parameters is a first step toward testing the model framework
at the LHC. Of course, it is equally important to quantify
the reach of the LHC in each of these channels which include,
\begin{itemize}
\item the non-leptonic $\eslt + \jets$ channel studied in Paper I,
\item the $1\ell + \jets + \eslt$ channel,
\item the opposite-sign (OS) $\dilepton + \jets +\eslt$ channel,
\item the same-sign (SS) $\dilepton + \jets +\eslt$ channel,
\item the ${\rm multilepton} + \jets + \eslt$ channel, with $n_{\ell}\geq 3$.
\end{itemize}
These signal channels ought to originate mainly from
squark and gluino pair production, followed by their cascade decays.
In addition, there are also clean ({\it i.e.} free from central jet activity)
channels with
\begin{itemize}
\item dilepton plus $\eslt$ events, and
\item trilepton and $n_{\ell} \geq 4$ lepton events,
\end{itemize}
mainly from the pair production of sleptons\cite{AMET,SLEP}
as well as from $\tw_1\tw_1$, $\tw_1\tz_2$\cite{BARB,TRILHC}
and $\tz_2\tz_2$\cite{BISSET} production processes.
Of course, after cuts, several sparticle
production mechanisms can contribute to each channel, so that
it is necessary to simultaneously generate the production
of {\it all} sparticles in order to obtain an accurate assessment
of the expected signals.

In this paper we continue the study of SUSY signals within the SUGRA
framework that we began in Paper I. We use
ISAJET 7.14\cite{ISAJET} to compute\cite{FN3} the signal cross sections
after cuts designed to separate the SUSY signals from SM backgrounds,
and wherever possible, also to separate SUSY sources from one another,
in each of the leptonic channels listed above.
We show these in the $m_0$ -- $\mhf$ plane, which provides a convenient
way to display the signals from different sparticle production processes.
For other parameters, our canonical choices are $A_0=0$, $\tan\beta=2$ and $10$
and we adopt both signs of $\mu$.
To orient the reader with various sparticles masses derived from the SUGRA
framework, we show contours of squark and gluino masses in
Fig.~\ref{SQGL}, and of slepton and chargino masses in Fig.~\ref{WINSL},
for ({\it a})~$\tan\beta=2,\mu<0$,
({\it b})~$\tan\beta=2,\mu>0$, ({\it c})~$\tan\beta=10,\mu<0$, and
({\it d})~$\tan\beta=10,\mu>0$. We remind the reader of the approximate
relationship $m_{\tz_2} \simeq m_{\tw_1} \simeq 2m_{\tz_1}$
that usually holds because $|\mu|$ tends to be large within this framework.
In Fig.~\ref{SQGL}, as well as in many subsequent figures,
the regions shaded by bricks (hatches) are excluded by theoretical
(experimental) constraints as discussed in Paper I.
The gluino mass contours in Fig.~\ref{SQGL}
are not exactly horizontal because of the difference\cite{POLE} between the
running and physical ({\it i.e.} pole) gluino mass.
In Fig.~\ref{WINSL}, we also show the region
where the ``spoiler" decay modes $\tz_2 \to \tz_1 H_{\ell}$ or
$\tz_2 \to \tz_1 Z$ are kinematically accessible (above the dotted contours);
in this region, leptonic decays of the $\tz_2$ are either very suppressed,
or have additional backgrounds from SM $Z$ boson production.

We map out the regions of parameter space
where these signals are observable at the LHC, and compare this
with the region that can be probed via the $\eslt$
channel\cite{BCPT,ATLAS} as delineated in Paper I. On the issue of the LHC
reach, our main new result is that the
$1\ell$ channel provides the greatest reach for supersymmetry.
However, the observation of signals in several channels is important,
since it can help to identify SUSY as the unique source of new physics.
We study jet and $B$-hadron multiplicity distributions,
as well as charge asymmetry distributions in the single lepton and
SS $\dilepton + \jets +\eslt$ channels, and dilepton flavor asymmetry
in the OS $\dilepton +\jets +\eslt$ channel as these can
provide information about the cascade decay chains of
gluinos and squarks\cite{BDKNT}.
We also identify regions of parameter space where the clean
trilepton and the clean OS, same-flavor dilepton signals are observable.
While these regions form a subset of the region where SUSY
may be probed via the multijet channels, an observation of these
signals will be important because they will signal
$\tw_1\tz_2$ and slepton production, respectively; {\it i.e.}
with suitable cuts described below, there is limited contamination
from other SUSY sources.

The rest of this paper is organized as follows. In the next Section,
we briefly discuss some computational details. Sec.~3 and Sec.~4
focus on the multilepton plus multijet and clean
multilepton channels, respectively. We present a comparative analysis of the
reach in various channels in Sec.~5,
and also consolidate the information about the underlying SUGRA
model parameters that might be obtained by studying distributions
in these various channels.

\section{Event Simulation}

We work within the framework of the minimal SUGRA model and
use ISAJET 7.14 to simulate the various leptonic signals for SUSY
listed above. The implementation of the SUGRA framework into
ISAJET has been described elsewhere\cite{BCMPT,BCPT} and will
not be repeated here. We generate {\it all} lowest order $2 \to 2$
SUSY subprocesses in our simulation of the multilepton plus multijet
signals
(except for $s$-channel Higgs boson mediated subprocesses). However,
for the simulation of the {\it clean} multilepton signals,
we have generated only slepton and chargino/neutralino events,
since gluino and squark decays will very seldom yield final states
without central jet activity\cite{FN4}.

For detector simulation at the LHC, we use the toy calorimeter
simulation package ISAPLT. We simulate calorimetry covering $-5<\eta
<5$ with cell size $\Delta\eta\times\Delta\phi =0.05\times 0.05$. We
take the hadronic energy resolution to be $50\% /\sqrt{E}\oplus 0.03$
for $|\eta |<3$, where $\oplus$ denotes addition in quadrature, and to
be $100\% /\sqrt{E}\oplus 0.07$ for $3<|\eta |<5$, to model the
effective $p_T$ resolution of the forward calorimeter including the
effects of shower spreading, which is otherwise neglected. We take
electromagnetic resolution to be $10\% /\sqrt{E}\oplus 0.01$.
Although we have included these resolutions, which are typical of
ATLAS\cite{ATLAS} and CMS\cite{CMS}, we have made no attempt to
estimate the effects of
cracks, edges, and other problem regions. Much more detailed detector
simulations are needed to understand the effects of such regions and
of the resulting non-Gaussian tails, particularly on the $\eslt$
resolution.

Jets are found using fixed cones of size $R=\sqrt{\Delta\eta^2
+\Delta\phi^2} =0.7$ using the ISAJET routine GETJET. Clusters with
$E_T>100$ GeV and $|\eta ({\rm jet})|<3$ are labeled as jets.
However, for the purpose of jet-veto only, clusters
with $E_T>25$~GeV and $|\eta ({\rm jet})|<3$ are regarded as jets.
Muons and electrons are classified as isolated if they have $p_T>10$ GeV,
$|\eta (\ell )|<2.5$, and the visible activity within a cone of $R
=0.3$ about the lepton direction is less than $E_T({\rm cone})=5$ GeV.

We assume an integrated luminosity of $10\,\fb^{-1}$, corresponding to
$10^{33}\,\cmsec$ for one year. Hence we feel justified in neglecting
the effects of pileup. We presume it would be possible to use the
maximum LHC luminosity, $10^{34}\,\cmsec$, to search for gluinos and
squarks with masses $\agt 1$ -- 2~TeV.

\section{Multilepton plus Multijet Signals for Supersymmetry}

\subsection{Classification of Signals and Event Selection}

For $m_{\tg},\ m_{\tq} \alt 1$~TeV, $\tg\tg$, $\tg\tq$ and $\tq\tq$
production is the dominant source of SUSY events at the LHC.
These production mechanisms, together with $\tg$ and $\tq$ cascade decays,
naturally lead to events with $n\,{\rm leptons} + m\,\jets +\eslt$, where
typically $n=0\hbox{ -- }4$ and $m \geq 2$.
These event topologies may also arise from
the production of gluinos and squarks in association with a chargino
or a neutralino. In addition,
direct production of charginos, neutralinos and sleptons
followed by cascade decays to $\tw_i$ or $\tz_j$ can lead to similar events.

Although in our simulation we generate all SUSY processes using
ISAJET, our cuts are designed to selectively pick out gluino and
squark events, whose characteristics are high transverse momentum jets
and large missing transverse energy. Furthermore, the $p_T$ of the
primary jets from gluinos, as well as the $\eslt$ are expected to scale
with $m_{\tg}$. In contrast, the momenta of leptons, produced far down
in the cascade decay chain from chargino and neutralino daughters,
will not scale in energy the same way as jets and $\eslt$ which can be
produced in the first step of the cascade decay.  Thus, following
Paper I, for the multilepton plus multijet signals for SUSY, we vary
the missing-energy and jet $E_T$ cuts using a parameter $E_T^c$ but
fix the lepton cuts:

\begin{itemize}
\item jet multiplicity, $n_{\rm jet} \geq 2$ (with $E_{T,{\rm jet}} >
100$~GeV),

\item transverse sphericity $S_T > 0.2$,

\item $E_T(j_1), \  E_T(j_2) \ > \ E_T^c$ and $\eslt > E_T^c$.
\end{itemize}

We classify the events by the multiplicity of {\it isolated}
leptons, and in the case
of dilepton events, we also distinguish between the OS and the SS
sample as these could have substantially different origins. For the
leptons we require,

\begin{itemize}
\item $p_T(\ell) > 20$~GeV ($\ell=e$ or $\mu$) and $M_T(\ell,\eslt) > 100$~GeV
for the $1\ell$ signal, and
\item $p_T(\ell_1,\ell_2) > 20$~GeV for $n=2,3,\ldots$ lepton signals. We
do not impose any $p_T(\ell)> E_T^c$ requirement on the leptons for
reasons explained above.
\end{itemize}

\subsection{Calculation of Backgrounds}

SM processes, particularly those involving the production of
heavy particles like the $W$ and $Z$ bosons, or the top quarks,
can mimic the leptonic signals listed above.
We have used ISAJET to evaluate the following SM backgrounds to these signals:
\begin{enumerate}
\item $t\bar{t}$ production, where the leptonic decays of the tops can
give up to two isolated leptons; for $n>2$ the additional lepton may
come from a $b$ or $c$ decay or from the fragmentation of additional jets
in the event, where the lepton is accidentally isolated;
\item $W$ and $Z$ boson + jet production, where additional jets and/or leptons
come from parton showering;
\item $WW$, $WZ$ and $ZZ$ production, where additional jets can again arise
from QCD radiation;
\item QCD jet production, where leptons can arise from decays of
heavy flavors produced directly or via gluon splitting.
\end{enumerate}
ISAJET includes higher order QCD and electroweak
effects in the branching approximation: {\it i.e.} it includes
quark and gluon as well as weak vector boson radiation, using exact
kinematics but only collinear dynamics. Thus, extra leptons can arise in any
of the above hard scattering subprocesses additionally from, for instance,
gluon splitting to top or bottom quark pairs, followed by their subsequent
decays, or by $W$ and $Z$ boson radiation.

The $E_T^c$ dependence of these background cross sections, obtained
using CTEQ2L parton distributions\cite{CTEQ},
is displayed in Fig.~\ref{BACK} for
({\it a})~$1\ell+\jets$ events, ({\it b})~OS $\dilepton +\jets$ events,
({\it c})~SS $\dilepton + \jets$ events, and ({\it b})~$3\ell+\jets$ events.
We see that $W+\jets$ production is generally the largest background,
except in the OS dilepton channel where the $t\bar{t}$ background
dominates for modest values of $E_T^c$.
Gauge boson pair production and QCD background sources
are essentially negligible, compared to backgrounds
from $W$, $Z$ and $t\bar{t}$ production.
The multilepton background from gauge boson pair production
is strongly suppressed, presumably because of the requirement
of two additional hard jets as well as $\eslt>E_T^c$.

The wiggles in the curves in Fig.~\ref{BACK} are a reflection of the
statistical
fluctuations in our simulation. We see that for modest values of $E_T^c$
the fluctuation in the biggest backgrounds are under control. We will
see later that in order to extract the reach we use $E_T^c =200$~GeV
in all but the $1\ell$ channel for which we use $E_T^c =400$~GeV. We
have checked that for these ranges we typically obtain at least
several tens of events passing the cuts in our simulation, so
that the statistical errors on the relevant background estimates
are in control. It should, of course, be remembered that our background
calculations are probably correct only to a factor $\sim 2$-3 due to the
inherent uncertainties associated with leading-log QCD, the
parton shower approximation, our idealistic detector simulation, {\it etc}.

In order to enable the reader to assess this
calculation, we have shown the details of the background calculation
in the various multilepton channels in Table~I for one
value of $E_T^c$. Since only a tiny
fraction of the events generated pass the cuts, it is necessary
to generate events in several ranges of hard scattering $p_T$ ($p_T^{HS}$)
for each SM process, and then combine these to obtain the background
cross section from each of these sources\cite{FN5}.
The results of our computation for $E_T^c=200$~GeV are shown in
Table~I,
for the $1\ell$, OS, SS and $3\ell$ signals. In those $p_T^{HS}$
bins where we obtain no events, the bound shown corresponds to the
cross section corresponding to the one-event level. We see that
for the $t\bar{t}$ and $W$ or $Z$ backgrounds which are the largest
contributors to the background cross section, the main contribution
indeed comes from the intermediate values of $p_T^{HS}$, ensuring
that we do have a reasonable estimate for the cross section. We have
also checked that for the major contributors to the background,
we have ten to several hundred events passing the cuts in
our simulation, so that our estimates should be reliable to
a few tens of percent, and frequently much better, as far as
statistical errors are concerned. Finally, we see that the
QCD background to the $1\ell$ cross section is clearly small;
while we typically obtain only a bound on this from our simulation,
it is reasonable to expect that this will not be a substantial
background in the multilepton channels.

We have also attempted to estimate the $4\ell$ background with ISAJET.
Such events, however, form an extremely tiny fraction of the total
cross section so that a reliable simulation of these would require
lengthy computer runs. Our simulation in which just a handful
of events pass the cuts in each of the $W+$jets, $Z+$jets and the
$t\bar{t}$ channels yields a cross section
$\sigma(4\ell)= 0.04\,\fb$ for this background for $E_T^c=100$~GeV,
which falls to 0.002~fb for $E_T^c=200$~GeV.
Even allowing for uncertainties in our estimates,
we see that the SM background is essentially negligible.
For reasons of brevity and because these signals are observable only
for limited ranges of parameters, we have not shown these cross
sections in the figures or in Table~I.

\subsection{SUSY Multilepton plus Multijet Signals at the LHC}

Our next goal is to evaluate the various SUSY
$n\hbox{-lepton}+m\hbox{-jets}+\eslt$ signals expected from
supersymmetry at the LHC, and compare against background expectations.
Toward this end, we show the signal cross sections along with the
total SM background as a function of $E_T^c$ in Fig.~\ref{SIGNAL} for
({\it a})~$1\ell+\jets$ events, ({\it b})~OS $\dilepton+\jets$
events, ({\it c})~SS $\dilepton+\jets$ events, and ({\it
b})~$3\ell+\jets$ events.  Our total signal and background cross
sections are evaluated, as usual, at leading-log level, and so are
uncertain to about a factor of 2; next-to-leading log gluino and
squark cross sections can be found in Ref.~\cite{ZERWAS}.  We have
illustrated the signal for the same six choices of SUSY parameters as
in Ref.~\cite{BCPT}; we take $A_0=0$, $\tan\beta=2$, $m_t=170$ GeV and
\begin{enumerate}
\item $m_0=\mhf =100$~GeV, for which $m_{\tg}=290$~GeV and $m_{\tq}=270$~GeV;
\item $m_0=4\mhf =400$~GeV, for which $m_{\tg}=310$~GeV and $m_{\tq}=460$~GeV;
\item $m_0=\mhf =300$~GeV, for which $m_{\tg}=770$~GeV and $m_{\tq}=720$~GeV;
\item $m_0=4\mhf =1200$~GeV, for which $m_{\tg}=830$~GeV and
$m_{\tq}=1350$~GeV;
\item $m_0=\mhf =600$~GeV, for which $m_{\tg}=1400$~GeV and $m_{\tq}=1300$~GeV;
\item $m_0=4\mhf =2000$~GeV, for which $m_{\tg}=1300$~GeV and
$m_{\tq}=2200$~GeV.
\end{enumerate}
{}From Fig.~\ref{SIGNAL} it is relatively obvious how $E_T^c$
should be chosen to search for SUSY in the multilepton plus multijet
channels: if gluinos are relatively light (cases 1 and 2),
$E_T^c\sim 100$ -- 150~GeV
suffices to obtain a large signal to background ratio and a large
event rate in all the channels. For the cases with heavier gluinos and
squarks (cases 3 -- 6), a larger value
of $E_T^c$ is necessary, though it should not be chosen too large
as to cut out all the signal. For instance, $E_T^c\sim 200$ GeV
should yield an observable signal, with a signal to background
ratio larger than unity in all but the OS dilepton channel. The {\it maximal}
reach may be
anticipated to occur in the $1\ell$ channel --- for cases 5 and 6,
with $E_T^c=400$~GeV, we expect $\sim 20$ -- 100 events (versus a background
of just about three or four events) after a year of LHC operation at
its ``low'' luminosity of $10\,\fb^{-1}/{\rm yr}$.

We next examine in detail each of the multilepton plus multijet topologies
as a function of SUGRA parameters.

\subsubsection{Single Lepton Events}

We begin by showing, in the $m_0$ -- $\mhf$ plane, cross section
contours for the $1\ell$ signal after the cuts discussed above for
$A_0=0$ and ({\it a})~$\tan\beta=2,\ \mu<0$, ({\it b})~$\tan\beta=2,\
\mu>0$, ({\it c})~$\tan\beta=10,\ \mu<0$, and ({\it
d})~$\tan\beta=10,\ \mu>0$ in Fig.~\ref{ONEL}. We have shown the
results for $E_T^c=100$~GeV (solid) for which the total SM background
from Fig.~\ref{SIGNAL} is $\sim 1300\,\fb$, and also for
$E_T^c=400$~GeV (dotted), for which the background is very tiny at
about 0.5~fb. For an integrated luminosity of $10\,\fb^{-1}$ the
corresponding $5\sigma$ limits are, 57~fb and 1.1~fb, respectively.

To obtain these contours (as well as the corresponding contours
for the OS, SS and $3\ell$ signals discussed below) we have first computed
the signal cross section for each point on a 100~GeV ~$\times$~100~GeV
lattice in the $m_0$ -- $\mhf$ plane, for points which do not fall inside
the excluded shaded regions. The contours are then obtained via
interpolation. We have cut off the contours near the boundaries of the
shaded regions where the sampling is poorer and the interpolation not as
reliable.

The curves shown are for cross sections of 1, 2, 4, 8, \ldots fb (only
every other solid curve is labeled) in each of these cases.
We have also checked that, even for the very hard $E_T^c$ cut, there are
sufficiently many events in our simulation to yield
reliable estimates of the cross sections: for the $E_T^c=400$~GeV
case, the efficiency for SUSY events to pass the cuts becomes very small
unless sparticles are rather heavy, so that for moderate
$\mhf$ values very lengthy computer runs would be necessary to
compute the cross section. For this reason, and because
very hard cuts are necessary only for the largest gluino and squark masses,
we have shown only the first three dotted curves, corresponding
to cross sections after cuts of 1, 2 and 4~fb.

We see from Fig.~\ref{ONEL} that  with $E_T^c= 100$~GeV, the $5\sigma$
reach (the 64~fb contour is closest to the 57~fb $5\sigma$ limit)
in the $1\ell$ channel extends to $\mhf\sim 600$~GeV for small values
of $m_0$ (corresponding to $m_{\tg} \sim m_{\tq} \sim 1.5$~TeV), or
to $\mhf \sim 400$~GeV ($m_{\tg} \sim 1$~TeV) if squarks are
heavy. Regions below this $5\sigma$ contour all have larger signal
cross sections, so that we found no ``holes'' of non-observability below
the $5\sigma$ limit. Notice, however, that the signal to background
ratio at the $5\sigma$ limit is just less than $5$\%: if we require this ratio
to exceed 25\%, the corresponding reach is between the 256 and 512~fb
contours.
To probe values of $\mhf \agt 250-300$~GeV, it is
best to choose larger values of $E_T^c$ to obtain better
statistical significance as well as higher signal to background ratio.
The maximal reach in the $1\ell$ channel can be obtained by using
a hard $E_T^c$ cut which eliminates essentially all the background
but still retains the signal at an observable level. The highest of the
dotted contours (1~fb) is very close to the $5\sigma$ limit for
$E_T^c=400$ GeV; in this case, ${\rm signal}/{\rm background} \sim 1$,
and $\mhf\sim 700$ GeV (1000 GeV) can be probed in the large (small)
$m_0$ region. This corresponds to a reach in $m_{\tg}\sim 1700$ GeV
(2300 GeV). Thus, we note that {\em the reach in
this channel appears to substantially exceed the corresponding
reach\cite{ATLAS,BCPT} in the canonical ${\it multijet}+\eslt$
(no isolated lepton) channel.}

How well can one determine the SUSY parameters by studying the
$1\ell +\jet+\eslt$ signal? Measurement of the total rate for such events
would localize, within errors, a position along one of the total
cross section contours of Fig.~\ref{ONEL}. These contours vary strongly
with $\mhf$, but less strongly with $m_0$. The cross sections
are roughly the same in all the four frames. While this means that
the 1$\ell$ signal rate yields no information about $\tan\beta$ or
$\sgn\mu$, it also means that a measurement --- and calculation --- of the
cross section
to within a factor of 2 would indeed tell us on which contour
we are within about $\pm 50$~GeV. While this does not accurately pin $\mhf$
because the contours are not quite horizontal, one would still be able to
obtain a reasonable estimate of $\mhf$.
The range of $\mhf$ thus
obtained can be checked for consistency with the gluino mass that might
be extracted\cite{BCPT}
from the $\eslt$ channel: in fact, a similar measurement ought to be possible
in the $1\ell$ channel.

Determination of $m_0$ is more difficult
and will probably require a simultaneous study of several signals
and their distributions. We note, however, that for very small
values of $m_0$, because of the enhancement of the leptonic decays of
$\tw_1$, and frequently also of $\tz_2$,
the lepton plus multijet cross sections are large. In contrast, the 0$\ell$
plus multijet cross sections (for a fixed value of $\mhf$) actually
reduce\cite{BCPT} as $m_0 \rightarrow 0$ because of the lepton veto: thus,
a measurement of the ratio of non-leptonic to multileptonic multijet
cross sections could yield information on whether we are in the
small $m_0$ region, particularly in the region where two body decays
of $\tw_1$ and/or $\tz_2$ into real sleptons are kinematically allowed
(this region has been delineated in Fig.~\ref{TRILEP} below).

Next, we turn to an examination
of the single lepton charge asymmetry, which could provide
an additional handle on position in parameter
space. Since
LHC is a $pp$ collider, there is a preponderance of valence $u$-quarks in the
initial scattering state, which can lead to a large proportion of $\tilde u$
squarks being produced in the final state if squarks are moderately
heavy and gluinos not too light (otherwise, sea parton annihilation
may be the dominant source of sparticles). The $\tilde u$-squarks frequently
decay to $\tw_1^+$, and one is led to expect more $\ell^+$s being produced
than $\ell^-$s. For large $m_0$ compared to $\mhf$, $\tg\tg$ production
should be dominant, which leads to equal production of $\ell^+$ and $\ell^-$
in cascade decays. Likewise, as we already noted,
if $\mhf$ is small, then sparticle production
at the LHC is dominated by gluon fusion and sea-quark annihilation, which
also leads to equal $\ell^+$ and $\ell^-$ production. In contrast, for
larger values of $\mhf$ and not too large $m_0$, the squark production via
valence quarks can dominate, and lead to the lepton charge asymmetry. To
illustrate this, we show in Fig.~\ref{L1AS} the single lepton charge
asymmetry
$$
A_c={N(\ell^+)-N(\ell^-)\over{N(\ell^+)+N(\ell^-)}}\,,
$$
versus $m_0$, for ({\it a}) $\mhf =100$ GeV, using $E_T^c=100$ GeV,
({\it b}) $\mhf =200$ GeV, using $E_T^c=200$ GeV, ({\it c})
$\mhf =400$ GeV, using $E_T^c=400$ GeV
and ({\it d}) $\mhf =500$ GeV, using $E_T^c=400$ GeV. In all frames,
$A_0=0$ and $\tan\beta =2$. For frames ({\it a} -- {\it d}), we take $\mu <0$;
frames ({\it e})-({\it h}) are the same except that $\mu >0$.
The horizontal dashed line is at $A_c=0$. The rather small SM background
(see Fig.~\ref{BACK}) has not been included in these figures.
We indeed see that for
the small $\mhf$ cases of frames ({\it a}) and ({\it e}), the asymmetry is
consistent with zero (by choosing a larger $E_T^c$ value, it may be
possible to enhance the valence contribution and so obtain an
asymmetry even in this case.). As we move up in $\mhf$ values, a significant
positive charge asymmetry develops, especially for small values of $m_0$,
reflecting the relative contribution of $\tilde u$-squarks versus
$\tilde d$-squarks, or other squark flavors or gluinos. Knowledge
of $\mhf$ may thus be combined with the measurement of $A_c$ to roughly
localize $m_0$ --- we would at least learn whether $m_0 \alt \mhf$ or
whether $m_0 \gg \mhf$.

In order to explore other strategies for the determination of $m_0$,
in Fig.~\ref{L1JM}
we have shown
the mean jet multiplicity ($\langle n_{j} \rangle$) as a function of
$m_0$ for the same eight cases ({\it a}) -- ({\it h}) as in Fig.~\ref{L1AS}.
We see that for a fixed value of $\mhf$ (which can be determined
from other considerations),
$\langle n_j \rangle$ clearly increases with $m_0$. The underlying
physics is exactly the same\cite{BCPT}
as for the $\eslt$ sample: for
small $m_0$, squark production is a significant source of $\eslt$
events, and because $\tq_R$ frequently directly decay via $\tq_R \to q \tz_1$,
the mean jet multiplicity is reduced. The mean jet multiplicity
is essentially independent of the sign of $\mu$. We see, however, that
it can increase by as much as a whole unit as $m_0$ varies between
100~GeV and 1~TeV. We also see that
the precision with which $m_0$ can be determined
depends on the values of other SUSY parameters. Finally, we note
that although $\langle n_{j} \rangle$ changes only by about 30\%
as $m_0$ is varied over the whole range in the figure, this could mean
a significant increase in the cross section for high multiplicity
(say $n_j \geq 4$ or 5) relative to $n_j=2$, so that ratios
of cross sections with different jet multiplicities could yield
a more sensitive measure of $m_0$. Devising the optimal
measure for localizing $m_0$ would require
a detailed study beyond the scope of the present analysis.

The multiplicity of tagged $B$-hadrons may also yield information
about the underlying parameters. Towards this end, in Fig.~\ref{NB1L}
we have plotted
the mean multiplicity ($\langle n_B \rangle$) of tagged $B$-hadrons
in the 1$\ell$ SUSY sample
for the same cases as in Fig.~\ref{L1JM}, assuming that
a $B$-hadron with $p_T > 20$~GeV and $|\eta_B |<2$
is tagged with an efficiency of 40\%.
We see from the figure that $\langle n_B \rangle$ varies
between 0.2 to 1.3 over the parameter
range shown. For the light gluinos cases
with $\mhf =100$~GeV in frames ({\it a}) and ({\it e}),
$\langle n_B \rangle$ is small and shows little variation with $m_0$, except
around $m_0=200$~GeV where the decays $\tg \to b\tb$ dominate other
squark decays. For
large $m_0$, the gluino decays via the three body modes, except that
the decays to tops (which can potentially be enhanced) are kinematically
suppressed. For heavier gluinos ({\it i.e.} larger values of $\mhf$) there
are two important differences. First, the spoiler decay
$\tz_2 \to \tz_1 H_{\ell}$, which is a source of $B$'s, may be kinematically
accessible: this leads to an increase
in $\langle n_B \rangle$ which is roughly independent
of $m_0$, except for the very small $m_0$ region
where neutralino decays to sleptons are also accessible.
Second, gluinos are heavy enough to decay to $t$ quarks. Thus,
when $m_0$ is very small, gluinos dominantly decay via $\tg \to q\tq$ into
all flavors. As $m_0$ is increased, $\tg\to\tst_1 t$ or $\tb_1 b$ may be
kinematically allowed, while the $\tg\to\tq q$ modes are closed.
For even larger values of $m_0$, three body decays to
third generation quarks can be enhanced due to propagator and
large Yukawa coupling effects\cite{BTWLOOP,BARTL},
leading to an increase in $\langle n_B \rangle$.
Finally, we remark that $\langle n_B \rangle$ does not serve to discriminate
between the two
signs of $\mu$. We caution the reader that $\langle n_B \rangle$
may potentially be sensitive to variations in $A_0$, since these may
alter the masses and mixings of third generation sfermions. Thus some
care must be exercised when attempting to extract $m_0$ from a measurement
of the $B$-hadron multiplicity.

\subsubsection{Opposite Sign Dilepton Events}

Cross section contours for the OS dilepton signal are shown in
Fig.~\ref{OSL} for the same cases as for the $1\ell$ signal in
Fig.~\ref{ONEL}. The solid lines are for $E_T^c=100$~GeV for which
the SM background is 630~fb, while the dashed lines are for
$E_T^c=200$~GeV for which the background is just 9~fb. As
in Fig.~\ref{ONEL} and in subsequent figures,
we show the dashed contours only for relatively large values
of $\mhf$ for which employing the larger $E_T^c$ value is really
essential. A striking feature of Fig.~\ref{OSL} is the
sharp kink near $m_0 \sim 400$~GeV where the contours
change their slope. In cases ({\it b}) -- ({\it d}) this is
simply due to the opening up of the two body decays of
the chargino and $\tz_2$ into $\tnu$ and $\tell_L$ (their branching
fractions to $\tell_R$ are strongly suppressed because
$\tz_2$ ($\tw_1$) has very small (zero) $U(1)$ gaugino components).
In case ({\it a}) however, the kink in the 1 and 2~fb contours
occurs at around $m_0=500$~GeV. We have checked that this is
because $\tz_2$ and $\tw_1$ leptonic {\it three-body} decays mediated
by left-handed sleptons
have significant branching fractions (few percent) even though the
two-body decays $\tz_2 \to Z\tz_1$ or $\tz_2\to H_{\ell}\tz_1$
and $\tw_1\to W\tz_1$
are kinematically accessible: the resulting enhancement of the
leptonic branching ratio, especially of $\tw_1$, accounts for the
kink being somewhat beyond the sleptonic two-body decay region in
case~({\it a}).

The 5$\sigma$ observability level is at 40~fb for $E_T^c=100$~GeV
and at 4.7~fb for $E_T^c=200$~GeV. We thus see from Fig.~\ref{OSL}
that with $E_T^c=100$~GeV,
the LHC should be able to observe a signal in this channel
if $\mhf \alt 300$ -- 400~GeV (200-300~GeV, if we also require
$\frac{S}{B}>0.25$).
The reach improves to $\mhf =$400-500~GeV if the analysis
is done using $E_T^c=200$~GeV. Notice that the reach is slightly
larger in the $\tan\beta=10$ cases than in the low $\tan\beta$ cases
({\it a}) and ({\it b}). This is because the branching fraction for the
two-body
$\tz_2 \to Z\tz_1$ decay, which is very small for
cases ({\it a}) and ({\it b}),
is sizable when $\tan\beta$ is large.
We have checked that the statistical
significance of the signal is
marginally improved with $E_T^c=300$~GeV, but the cross section
is then just around 1~fb for $\mhf =500$~GeV. In summary,
with suitable cuts and $10\,\fb^{-1}$ of data, LHC experiments
should be able to detect a signal in the OS dilepton channel
for $\mhf$ up to 400-500~GeV, which corresponds to a gluino mass just beyond
1~TeV.

If an OS dilepton signal is seen, one may again attempt to localize
the position in parameter space via a measurement of the total OS
dilepton cross section, which should place one along one of the
contours in Fig.~\ref{OSL}. Since the $\tz_2$ branching ratio into $Z$
bosons depends on $\tan\beta$, the number of reconstructed
$Z\to\ell\bar{\ell}$ events may offer some rough discrimination in that
parameter if the spoiler modes are kinematically accessible. An idea
of $\mhf$ from the $\eslt$ or $1\ell$ channels together with the cross
section in this channel would enable the determination of $m_0$ if it
is small: for example, for $\mhf =400$~GeV, the OS cross section
rapidly varies from $>128\,\fb$ (small $m_0$) down to 32~fb ($m_0 \sim
500$~GeV), and then slowly decreases to $\alt 8\,\fb$. Ratios such as
${\sigma (OS)}/{\sigma(1\ell)}$ or ${\sigma (OS)}/{\sigma(0\ell)}$
would presumably be more accurately calculable than the absolute cross
sections.

It was noted in Ref.~\cite{BDKNT} that the production of
neutralinos in SUSY events can lead to a flavor asymmetry
in the OS dilepton event sample, which may allow further
parameter space location. For instance, if OS dileptons
are primarily coming from $\tz_2$ decay, then they should mainly be of
same flavor, {\it e.g.} $e\bar{e}$ or $\mu\bar{\mu}$ pairs. If instead OS
dileptons come mainly from charginos or third generation
quarks and squarks and their subsequent leptonic
decays, then one would expect roughly equal abundance of
$e\bar{\mu}$ and $\mu\bar{e}$ pairs as compared to same flavor
lepton pairs. We have plotted the OS dilepton
flavor asymmetry,
$$
A_F={{N(e\bar{e})+N(\mu\bar{\mu})-N(e\bar{\mu})-N(\mu\bar{e})}\over
{N(e\bar{e})+N(\mu\bar{\mu})+N(e\bar{\mu})+N(\mu\bar{e})})}\,.
$$
in Fig.~\ref{FLAS} for the same cases as in the previous figure.  SM
backgrounds are included in the figure.  Points denoted by an x have
an asymmetry $A_F<0.2$, consistent with no asymmetry in our
simulation.  Open boxes or diamonds have asymmetry $0.2<A_F<0.5$, while
filled boxes or diamonds have $A_F>0.5$.  For $\mhf =100$~GeV, we took
$E_T^c=100$ GeV, and we use the box symbols.  For larger $\mhf$, we
took $E_T^c=200$~GeV to improve the signal/background, and we use the
diamond symbols.  We see in frame ({\it a}) that there is a large
asymmetry for $\mhf\alt 200$ GeV, and also for small $m_0$ values. In
the former case, for small $m_0$, this is due to an enhanced branching
fraction for $\tz_2\to\ell\bar{\ell}\tz_1$, while in the latter case
it is due in part to the $\tz_2\to\ell\tell_i$ two-body decays. In
frame ({\it b}), the asymmetry disappears for small $\mhf$ values due
to interference effects driving the $\tz_2$ branching fraction to very
small values\cite{TEVSTAR,MRENNA}. The two frames for $\tan\beta =10$
continue to have significant flavor asymmetry even for $\mhf$ as high
as $\sim 300$ GeV due to the significant $\tz_2\to Z\tz_1$ branching
fraction. Again the significant asymmetry for small values of $m_0$ is
due to real slepton decays of the neutralino. Note also that in some
cases there is an observable asymmetry even when the leptonic
branching fraction of $\tz_2$ is so small that the clean $3\ell$ signal
from $\tw_1\tz_2$ production (discussed in Sec.~IV) falls below the
observable level. We thus see that an observation of a significant
flavor asymmetry will localize us in the regions of the plane where at
least one of $m_0$ or $\mhf$ is not too large. Furthermore, if event
rates indicate a large value of $\mhf$, the observation of a flavor
asymmetry would lead us to conclude that $m_0$ is rather small.

We also mention that we have checked that the jet multiplicity increases
with $m_0$ for $\mhf =200$ and 400~GeV, with other parameters fixed
as in Fig.~\ref{L1JM}. We have also checked the $\langle n_B \rangle$
distributions
which show a qualitatively similar trend as in
the $1\ell$ case shown above. Again the results are essentially the same for
the two signs of $\mu$. We do not show these distributions for the sake
of brevity.

\subsubsection{Same Sign Dilepton Events}

The SS dilepton plus jets channel has long been known\cite{BKP,BGH,EARLY} to
provide a clean signature for supersymmetry, and has been the subject
of several studies\cite{EARLY,ATLAS,BGH,BDKNT,DP} at the LHC. Within the MSSM
framework, detailed simulations\cite{ATLAS} have shown that,
in this channel, the gluino
reach extends to beyond 1~TeV. It is also known\cite{BDKNT,DP} that while
gluino pair production with gluinos decaying via
the chain $\tg \to \tw_1 \to \ell$ is frequently considered to be
the main source of these events, many other sources may be important, in
particular, decays of gluinos to third generation fermions and also squark
decays.
We also stress that same sign dilepton
events do not necessarily originate via production
of Majorana particles. For instance the production of $\tb_1\bar{\tb}_1$
pairs, where $\tb_1 \to t\tw_1$ can also lead to SS dilepton plus
multijet topologies.
It is clear that a reliable computation of the SS dilepton signal
requires that all the decay chains as well as all possible production
mechanisms be included, as is done in ISAJET.

The SS dilepton cross section is shown in Fig.~\ref{SSL} for the
same values of SUGRA parameters as in Fig.~\ref{ONEL} for $E_T^c = 100$~GeV
(solid contours) and $E_T^c = 200$~GeV (dashed contours).
As in Fig.~\ref{OSL} (and for essentially the same reasons) we
see that the contours show a kink near the region where $\tell_L$
and $\tnu$ masses approach $m_{\tw_1} \simeq m_{\tz_2}$.
The SM backgrounds
to the signal are just 1.7~fb and 0.25~fb, respectively, yielding
``5$\sigma$'' limits of 2.1~fb and 0.8~fb. (For $E_T^c=200$~GeV,
the Poisson probability
of an expected background of 2.5 events fluctuating to 8 events
is $4\times 10^{-3}$, so that the 10 event level is perhaps a more
reasonable estimate of the reach.) We see from Fig.~\ref{SSL} that
even with $E_T^c=100$~GeV, the $5\sigma$ reach extends out to
$\mhf\sim 400$ -- 500~GeV, and up to 700~GeV in the small $m_0$ region,
where squarks
are relatively light and leptonic decays of $\tw_1$ and $\tz_2$ are
enhanced.  The signal to background ratio
exceeds unity. A higher value of $E_T^c$ only gives a marginal
increase in the reach. With the harder cut, the signal is small
so that perhaps $20\,\fb^{-1}$ of integrated luminosity may be
necessary in this case. As in Fig.~\ref{OSL}, the cross sections
are somewhat larger in the $\tan\beta=10$ cases as compared to the
$\tan\beta=2$ cases.
Finally, we note that although there are some fluctuations in our
simulation for $\mhf < 200$~GeV, this signal should again be observable
down to relatively low values of $\mhf$.

Again, a measurement of the total cross section for SS dilepton
events will place us along one of the contours in the $m_0$ -- $\mhf$ plane.
As before, a measurement of the ratio of the SS cross section to the
$1\ell$ cross section would be an indicator of the small $m_0$, large
$\mhf$ region. As in the OS dilepton case, the cross sections
are somewhat larger for the high $\tan\beta$ cases.
It has been pointed out\cite{EARLY} that a charge
asymmetry may exist in any SS dilepton signal detected at a $pp$
collider; this observation has since been confirmed by more
detailed simulations\cite{ATLAS}.
As for the asymmetry in the $1\ell$ channel, the SS dilepton charge
asymmetry is again a reflection of the valence $u$ and $d$ quarks in the
proton participating in the production mechanism. We show in Fig.~\ref{SSAS}
the charge asymmetry,
$$
A_c={{N(\ell^+\ell^+)-N(\ell^-\ell^-)}\over
{N(\ell^+\ell^+)+N(\ell^-\ell^-)}}\,,
$$
as a function of $m_0$, for ({\it a}) $\mhf =200$ GeV (with $E_T^c=100$ GeV),
and ({\it b}) $\mhf =400$ GeV (with $E_T^c=200$ GeV). We take
$A_0=0$, $\tan\beta =2$ and $\mu <0$. Frames ({\it c})
and ({\it d}) are same as ({\it a}) and ({\it b}),
except for the sign of $\mu$. It can be noted that for
$m_0$ large, where $\tg\tg$ production is dominant, there is essentially
no charge asymmetry. As $m_0$ decreases, and squarks become lighter,
the asymmetry grows, reflecting the presence of $\tilde u$-squarks as a
significant
source of the event sample.

We have checked the dependence of the jet multiplicity in
this sample on $m_0$ for the same cases as in Fig.~\ref{SSAS}. While
we do see the anticipated trend for an increase of $\langle n_{j} \rangle$
with $m_0$, the extraction of $m_0$ appears more difficult than in the
$1\ell$ case, in part because of somewhat larger error bars in our
simulation. We have also checked the $\langle n_B \rangle$ distributions for
these
same cases  --- they appear
to be qualitatively similar to those for the $1\ell$ and OS dilepton samples.

Finally, we have checked the sources of SS dilepton events for
several points in parameter space to see how frequently these
occur when each gluino decays via $\tg \to q\bar{q}\tw_1$ (where $q \not = t$)
and $\tw_1$ decays leptonically since this chain has been suggested\cite{BGH}
as a way for extracting $m_{\tg}$. For small values of $m_0 \simeq \mhf$
this does not happen because gluinos decay to real squarks. For
$\mhf =250$~GeV and $m_0=4\mhf$,
we have checked the sources in cases ({\it a}) -- ({\it c}) of Fig.~\ref{SSL}.
We found that in these three cases, just 2/51, 3/36 and 0/48 events
came from this particular cascade chain.
Typically, in about half the SS dilepton event
sample that passed the cuts, there was at least one $t$-quark from
the decay of the gluino (recall that decays to third generation
may be enhanced), and greater than a third of the events
had their origin in $\tg\tq_L$ or $\tg\tq_R$ production even though
the squarks were somewhat heavier than gluinos. While these numbers
do depend on the details of the cuts, they underscore the importance
of simulating all possible production mechanisms and decay chains
in order to make a realistic assessment of the feasibility of mass measurement
in this channel.

\subsubsection{Trilepton Events}

Finally, we show the cross section contours for $3\ell + \jets +\eslt$
events in Fig.~\ref{TRIL}, again for the same four cases ({\it a}) -- ({\it
d});
as before, $E_T^c=100$~GeV (solid) and 200~GeV (dashed). We
note the following:
\begin{itemize}
\item As expected, the cross sections are enhanced in the region
where the two body decays of $\tw_1$ and $\tz_2$ to $\tell_L$ and $\tnu$
are kinematically allowed.

\item The cross sections remain substantial
even in the region of the plane where the spoiler decays
of $\tz_2$ (the boundaries of these regions are denoted by dotted lines in
Fig.~\ref{WINSL}) become kinematically allowed. We thus conclude that
while the cascade decay chains
$\tg\to q\bar{q}\tz_2\to q\bar{q} \ell\bar{\ell}\tz_1$ or
$\tq \to q \tz_2  \to q \ell\bar{\ell}\tz_1$ are important sources
of leptons in these events, there must be other sources operating as well.
These include cascade decays of squarks and gluinos
to charginos (which decay to $W$ bosons) and to top quarks.
We remind the reader that gluino decays to third generation
quarks can frequently be enhanced\cite{BTWLOOP,BARTL,DP}
because $\tt_1$ and $\tb_1$ are frequently
lighter than other squarks (so there is more phase space), because the third
generation Yukawa couplings can be large and enhance such decays,
and because Higgs bosons that are produced in the decay cascades
decay either to third generation fermions, or to charginos and/or neutralinos
(which have enhanced decays to third generation fermions).

\item For $\mhf \agt 250$~GeV (the boundary
of the $Z$ spoiler in Fig.~\ref{WINSL}) some of the trilepton sample
(or the dilepton sample in Fig.~\ref{OSL}) should consist of real $Z+\ell$
(real $Z$) events.

\item There are regions of parameter space \cite{TEVSTAR,MRENNA} where
the leptonic decays of $\tz_2$, and hence the $3\ell$ signal, are suppressed.
This causes a dip in the cross sections of frames
({\it c}) and ({\it d}) around $(m_0,\mhf ) \sim (300,200)$~GeV.
It is instructive to compare this with the corresponding case for the
clean trilepton signal from $\tw_1\tz_2\to 3\ell$ discussed
in the next Section.
\end{itemize}

We find
background cross sections of 4~fb and 0.07~fb, respectively, for
the two choices of $E_T^c$. The $5\sigma$ level for $E_T^c$
corresponds to a cross section of 3~fb, while the 5 -- 10 event
level might be a reasonable estimate for the reach with the
larger value of $E_T^c$. We thus see that with $E_T^c =100$~GeV,
the reach in the trilepton channel extends up to 350-500~GeV depending
on the parameters, except of course in the small $m_0$ region where
$\mhf$ values as high as 700~GeV may be probed. Again, the reach is
larger in the $\tan\beta=10$~cases. With $E_T^c=200$~GeV, the reach
in $\mhf$ increases by about 50 GeV. However, since the background
is essentially negligible in this case, it may be possible to push the limits
even further with a larger data sample.

As with the other multilepton channels discussed earlier, there should
be distinctive properties of the $3\ell+\jets+\eslt$ signal that allow
some localization of where one is in parameter space. These include
jet multiplicity, $B$ multiplicity and various $p_T$ distributions. Also,
there should again be a charge asymmetry, where we expect more $++-$
events than $+--$. We have also checked the variation of
$\langle n_{j}\rangle$
with $m_0$. The distributions are somewhat flatter in this case
as compared to the $1\ell$ sample in Fig.~\ref{L1JM}. This is
reasonable since for higher lepton multiplicity,
the number of ``partonic jets'' must be correspondingly reduced. In contrast,
the $\langle n_B \rangle$ distributions are qualitatively
similar to the previous cases.
We do not present these plots here for brevity.

\subsubsection{A recapitulation of the LHC reach via multijet plus
multilepton events}

We have seen that, regardless of the model parameters, for $m_{\tg}
\alt 1$~TeV ($\mhf \alt 400$~GeV) there should be an observable SUSY
signal in each of the $1\ell$, OS, SS and $3\ell$ channels if the
SUGRA framework that we have adopted is a reasonable description of
nature. In our previous study\cite{BCPT}, we saw that there will also
be a clearly observable signal in the $\jets+\eslt$ channel.  Hence,
in SUGRA, a wide variety of supersymmetric signals are expected to
occur at the LHC.  If gluinos are heavier than 1~TeV, the signals in
the di- and tri-lepton channels may not be observable, although
signals in the multijet$+\eslt$ and $1\ell$ channels may still be
visible.  The single lepton channel yields the maximal reach.  Our
computation shows that at the LHC experiments will probe $\mhf$ up to
600 -- 700~GeV ($m_{\tg}$ up to 1500 -- 1800~GeV) even if $m_0$ is
very large; if $m_0$ is relatively small, it will be possible to
search for gluinos heavier than 2~TeV \cite{FN6}.

Before closing this discussion, we should also mention that we have
examined the $4\ell$ channel\cite{EARLY}. We find that, with $E_T^c=100$~GeV,
these signals might be observable when $\mhf \alt 300$~GeV for $\tan\beta=2$
(or 500~GeV for $\tan\beta=10$), the reach in this channel is
always smaller than in other channels. For this reason, and because
there are rather few events in our simulation, we do not show
these here. It should, however, be kept in mind that sparticle
production can lead to these striking events at an observable level,
especially if $m_0$ is small and $\mhf$ not very large. Reducing
or even eliminating the jet cuts could lead to larger signals
in these event topologies without large increase in the background
(assuming that leptonically decaying $Z$ bosons can be readily identified).
We do not consider this any further in this study.

Up to now, we have fixed $A_0=0$ in our analysis.
The cross sections should mainly depend on $A_0$ due to
the variation of third generation squark masses.
Instead of performing lengthy scans of the parameter space, we have illustrated
the $A_0$ dependence of the cross section in Fig.~\ref{AZERO} for six choices
of $A_0$ and for
$m_0=500$~GeV, $\mhf=160$~GeV, $\tan\beta=2$ and $\mu >0$.
For $A_0=0,$ 500 and 1000 GeV, gluinos decay via three-body modes into
quarks plus various charginos and neutralinos. For larger, positive values
of $A_0$, the decay patterns of $\tt_1$ are qualitatively similar to
those for $A_0=1000$~GeV until $A_0$ exceeds $\sim 1330$~GeV, at which point
$m_{\tt_R}^2$ becomes negative.
For the three negative
values of $A_0$ sampled, the top-squark is so light that $\tg\to t\tst_1$
dominates the gluino decay channels (this is sensitive
to other model parameters, including $\sgn \ \mu$).
For $A_0=-500$ and $-900$ GeV, the $\tt_1 $ dominantly decays via
$\tst_1\to b\tw_1$, while for $A_0=-950$ GeV, it
is so light that only $\tst_1\to c\tz_1$ is allowed.
For all $A_0$ values sampled, we see that the $1\ell$ cross section is roughly
constant to within a factor of $\leq 2$.
The dilepton cross sections show a somewhat larger variation, although
this may not be sufficient, by itself, to determine $A_0$, since
small changes in $\mhf$ can cause similar variation. The maximum
variation is seen in the $3\ell$ cross section. The dilepton and
trilepton cross sections are largest for cases with large negative $A_0$
values of for which
$\tg \to t\bar{\tt}_1$ and $\tg \to \bar{t}\tt_1$ are the only
two body decays of the gluino. Since $\tt_1$ decays via $\tt_1\to b\tw_1$,
the increase in the leptonic cross sections (particularly for the $3\ell$
channel) should not be surprising. The sharp drop in the cross sections
at the most negative value of $A_0$
is because the chargino decay mode of the $t$-squark (which
is a source of leptons) becomes inaccessible,
and $\tt_1 \to c\tz_1$. In this case, $\tg\tg$ pairs (with $\tg \to t\tt_1$)
can give rise to events with at most two hard, isolated leptons, and
$\sigma(SS) \simeq \sigma(OS)$.
It may ultimately be possible from the ratios of multilepton to single
lepton cross sections to pin down $A_0$, especially if it is
close to the boundary of the excluded region where $m_{\tt_1}^2$
becomes negative. We have, however, seen
that this ratio shows a similar trend in the small $m_0$ region
where $\tw_1$ and $\tz_2$ leptonic decays are enhanced.
A measurement of $\langle n_B \rangle$ could serve to distinguish
the two different origins of leptonic signals. We also note that
in principle, there could be parameter
values for which $\tg \to b\tb_1$ might be the only allowed two body
gluino decay, in which case we would expect a reduction of
the multilepton cross sections.
More detailed study of the variation of the signals with $A_0$ are
clearly necessary before definitive conclusions can be drawn.

Our preliminary conclusions based on Fig.~\ref{AZERO} are
that {\it i}) the multi-lepton cross sections examined above
are less sensitive to variation in $A_0$ for lower lepton
multiplicity, and {\it ii})
except for the extreme cases where new channels for gluino
decays open up (these might be signaled by events with unusually
high $B$-hadron multiplicity), even
the multilepton cross sections are rather insensitive to $A_0$,
and the choice $A_0=0$ that we have adopted yields representative
values of these cross sections.

\section{Clean Multilepton Signatures for Supersymmetry}

In the previous section, we focussed on the study of
multilepton events with at least two hard jets and substantial
$\eslt$. The cascade decays of gluinos and squarks were the main
source of these jetty events.
While the direct production of charginos, neutralinos and sleptons
can also lead to similar event topologies,
these signals would be more difficult to pick out from SM
backgrounds because of relatively lower total cross sections
and softer $p_T(jet)$ and $\eslt$ distributions.
Moreover, there would be the additional issue of how to separate them from
the corresponding signals from gluino
and squark cascades for which the cross sections are considerably
larger. Clean multilepton events, {\it i.e.} events without any
jet activity, for which SM backgrounds are smaller, offer
a more promising way of searching for chargino and neutralino\cite{BARB,TRILHC}
or slepton\cite{AMET,SLEP} signals at the LHC. We study the
reach in SUGRA parameter space in these channels in this section.

Unlike in the previous section, where for each point in SUGRA parameter
space we generated {\it all} SUSY subprocesses using ISAJET, here we focus on
specific sets of reactions. This is because the majority of events
generated contain gluinos and squarks which almost always yield hard
jets, so that the efficiency for generating {\it clean} multilepton events
is very small: the computer time that would be necessary
to obtain an adequate sample of clean multilepton events would then make
global scans of SUGRA space quite intractable. For sample points
in the parameter space, we have checked how various SUSY channels
contribute to the specific reactions that we are searching for.

\subsection{Clean Trilepton Events from $\tw_1\tz_2$ Production}

These signals have previously been studied\cite{BARB,TRILHC}
within the framework of the MSSM for parameter sets motivated by SUGRA models.
In our previous study\cite{TRILHC}, we had
fixed $\mu =-m_{\tg}$ and chosen $\tan\beta=2$ and $m_{\tq}=m_{\tg}+20$~GeV.
We found that it was possible to find cuts which not only reduce SM backgrounds
to negligible levels, but also isolate trileptons produced via
$pp \to \tw_1\tz_2 + X \to \ell\nu\tz_1 + \ell'\bar{\ell'}\tz_1 + X$
from those produced by other SUSY reactions. These other SUSY
processes typically contribute
$\alt 10$\% of the total trilepton signal, at least for the parameters
where the signal was deemed to be observable.
Here, we extend our previous study and explore the reach of the
LHC for this signal within the SUGRA framework, and delineate the region
of parameter space where the clean trilepton signal should be observable
above SM backgrounds.

Exactly as in Ref.\cite{TRILHC}, we require:
\begin{itemize}
\item {\it i}) three isolated leptons, with $p_T(\ell_1,\ell_2) > 20$~GeV,
$p_T(\ell_3) > 10$~GeV.
\item {\it ii}) a central jet veto, {\it i.e.} no jet with
$p_T(jet)>25$~GeV within $|\eta_j|<3$;
\item {\it iii}) $\eslt < 100$~GeV;
\item {\it iv}) $|m(\ell \bar{\ell})-M_Z|>8$~GeV for all combinations of
OS leptons with the same flavor in the trilepton event.
\end{itemize}
Cuts {\it ii}) and {\it iii}) greatly reduce the backgrounds from
the cascade decays of gluinos and squarks, while {\it iv}) is designed
to eliminate $WZ$ events. After these cuts $t\bar{t}$ remains the dominant
background. It can be greatly reduced by further requiring,
\begin{itemize}
\item {\it v}) the two fastest leptons have the same sign of charge and the
flavor of
the slow lepton be the anti-flavor of either of the two fast leptons.
\end{itemize}
This reduces the signal by 50\% but essentially eliminates the top background,
from which the two hardest leptons almost always come from the primary decays
of the $t$-quarks, and hence, have opposite signs of charge. To recover
some of the
rejected signal without a significant increase in the $t\bar{t}$ background,
\begin{itemize}
\item {\it vi}) we retain events in which
the two fastest leptons have opposite sign provided $p_T(\ell_3)>20$~GeV.
\end{itemize}

After cuts {\it i}) -- {\it iv}) and either {\it v}) or {\it vi}), we
find a SM background level\cite{PHD} of 0.7~fb from $WZ$ production
where the gauge bosons decay into $e$, $\mu$ or $\tau$ (which then
decays leptonically), and 0.13~fb from $t\bar{t}$ production (for
$m_t=170$~GeV), yielding a total SM background of 0.83~fb. This is
somewhat larger than in our earlier study\cite{TRILHC} because of
differences in parton distributions as well as calorimeter simulation.
Assuming an integrated luminosity of $10\,\fb^{-1}$, the minimum
signal cross section for observability at the ``5$\sigma$ level"
($N_{\rm signal}>5\sqrt{N_{\rm bkgds}}$) works out to be 1.44~fb: the
Poisson probability of an upward fluctuation of this amount is
2$\times 10^{-5}$. Notice that $N_{\rm signal}/N_{\rm bkgd} \geq 1.7$.

The region of the $m_0$ -- $\mhf$ plane where the signal is observable
at the $5\sigma$ (10$\sigma$) level is shown by hollow (solid) squares
in Fig.~\ref{TRILEP} for $A_0=0$ and ({\it a})~$\tan\beta=2,\mu<0$,
({\it b})~$\tan\beta=2,\mu>0$, ({\it c})~$\tan\beta=10,\mu<0$, and
({\it d})~$\tan\beta=10,\mu>0$. For each parameter space point
sampled, we require at
least 25 events to pass the cuts in our simulation.
The x's show the points that we
have sampled but for which the signal falls below the $5\sigma$ level.
Also shown in Fig.~\ref{TRILEP} are the boundaries
of the region where the spoiler modes $\tz_2 \to \tz_1 H_{\ell}$ or
$\tz_2 \to \tz_1 Z$, or two body lepton-slepton decays of the neutralino
become accessible. In cases {\it b}) -- {\it d}) the boundary of the Higgs
spoiler decay is not shown as it always lies above the boundary of the $Z$
spoiler. Several features of this figure are worthy of mention:
\begin{itemize}
\item In case {\it a}), which corresponds most closely to the points sampled
in Ref.\cite{TRILHC}, we see that the signal is observable
at the $5\sigma$ level all the way up to the boundary of the
spoiler modes, and for most of the region the significance is
larger than $10\sigma$.

\item There are regions of the $m_0$ -- $\mhf$ plane in cases {\it b})
-- {\it d}) where the chargino is at its current experimental bound from LEP,
but where the trilepton signal fails to satisfy our 5$\sigma$ criterion
for observability. This was traced\cite{BT,TEVSTAR,MRENNA} directly
to the leptonic branching fraction of $\tz_2$ which can drop by
as much as two orders of magnitude due to interference
effects between the slepton and $Z$ mediated decay amplitudes.
Thus a non-observation of a signal in this channel will not
allow us to infer a lower limit on either $m_{\tw_1}$ or $m_{\tz_2}$.
The regions of the parameter plane where there is an observable
signal in this channel at the LHC are similar to
the regions that the Tevatron operating at
$10^{33}\,\cmsec$ could probe\cite{TEVSTAR}.

\item Except in the ``hole" mentioned above where there is no observable
signal,
the trilepton signal should be detectable all the way up to the limit
of the spoilers.
If sleptons are light enough so that
$\tz_2 \to \tell_{L,R}\ell$ are kinematically accessible (the small
$m_0$ region of the plane), then
these decays may dominate the spoiler decays. Then the branching fraction
for leptonic decays of $\tz_2$ is very large, and the reach in the
trilepton channel extends well beyond the boundary where the spoilers
become accessible. Notice the small wedge between the contours
labeled $\tz_2\to \tell_L\ell$ and $\tz_2 \to \tnu\nu$ where the
signal drops because the invisible decay $\tz_2 \to \tnu\nu$ of
the neutralino dominates.

\item We see that flipping the sign of $\mu$ makes a much larger difference
in the $\tan\beta=2$ cases
{\it a}) and {\it b}) relative to the $\tan\beta=10$ cases
{\it c}) and {\it d}). This can be understood if we recall that it is
always possible to choose $\mu$ and the gaugino masses to be
positive by convention: then,
the vacuum expectation values of the two Higgs fields can no longer be
chosen to be always positive, and the physically relevant sign
between $\mu$ and the gaugino masses appears as the sign of $\tan\beta$.
Of course, for large values of $\tan\beta$ where one of the vacuum
expectation values is essentially negligible, this sign is unimportant,
explaining why the results in cases {\it c}) and {\it d}) appear
so similar.

\end{itemize}

MSSM case studies of Ref.\cite{TRILHC}
suggest that the clean trilepton signal is relatively pure, and
that the ``contamination" from SUSY sources other than $\tw_1\tz_2$
production is small. It should be kept in mind that in these studies
we had fixed $\tan\beta=2$ and chosen $\mu=-m_{\tg}$, so that the
situation is roughly that in Fig.~\ref{TRILEP}{\it a}, where the
signal exceeds 10$\sigma$ over most of the plane. There are substantial
regions of the parameter plane in cases {\it b}) -- {\it d}) where
the significance of the signal is between 5 and 10$\sigma$.

It has already been pointed out\cite{TRILHC} that the isolation of
the signal from $\tw_1\tz_2$ production will allow a reliable
determination of $m_{\tz_2}-m_{\tz_1}$, and perhaps also other
combinations of chargino and neutralino masses.

\subsection{Clean Dilepton Events from $\tw_1\overline{\tw}_1$ Production}

We have just seen that while charginos and neutralinos might be detectable
over a large regions of parameter space in the clean trilepton channel,
there are parameter ranges for which the leptonic decays of $\tz_2$, and hence
this signal, is strongly suppressed even if charginos are relatively
light. We are thus led to examine whether OS dilepton signals
from the reaction $pp \to \tw_1\overline{\tw}_1+X \to \ell\bar{\nu}\tz_1+
\bar{\ell'}\nu\tz_1+X$ might be able to probe charginos in these regions,
or to provide a new channel for confirmation of the existence of
charginos detected in the trilepton channel; this was found to (at least
partially) be the case for the $10^{33}\,\cmsec$ upgrade of the
Tevatron\cite{TEVSTAR}.

To search for events in this channel we have made the following cuts:
\begin{itemize}
\item We focus on $e^{\pm}\mu^{\mp}$ events with $|p_T(\ell)|>20$~GeV
to eliminate large backgrounds from Drell-Yan production;
\item We veto events with any jet with $E_T>25$~GeV within $|\eta|<3$;
\item We require $30^0<\Delta\phi_{e\mu}<150^0$;
\item We require $\Delta\phi(\vec{p_T}(e\mu),\vec{\eslt})>160^0$;
\item We require 40~GeV~$<\eslt<100$~GeV (the upper limit on $\eslt$
is to prevent other SUSY sources from contaminating the signal).
\end{itemize}

We have used ISAJET to compute SM backgrounds to the dilepton signal
from $t\bar{t}$, $WW$, $\tau\bar{\tau}$, $WZ$ and $ZZ$ production.
We find that our
cuts efficiently suppress backgrounds from all but $WW$
events, for which the cut cross section is 136~fb (compared
to the $\sigma(t\bar{t})=9.9\,\fb$ and $\sigma(\tau\bar{\tau})=1\,\fb$).
For an integrated luminosity of $10\,\fb^{-1}$, the 5$\sigma$ level
of observability corresponds to a signal cross section of 19~fb,
although the signal/background ratio is small.
We sampled points in the $m_0$ -- $\mhf$ plane for the same cases
as for the clean trilepton signal in the previous subsection.
We found that except for a few points near $m_0=0$ in case~{\it a})
and an isolated point in case~{\it c}), the signal is below
the $5\sigma$ level, and for most of the plane, even below
the 3$\sigma$ level. We conclude that, unlike at Tevatron upgrades,
the dilepton signal from chargino pair production is unlikely to be
observable above SM backgrounds.

\subsection{Clean Dilepton Signals from Slepton Pair Production}

Charged sleptons and sneutrinos can be pair produced at the LHC in
$q\bar{q}$ fusion processes via charged or neutral gauge boson
exchange in the $s$-channel. Their (cascade) decays can lead to event
topologies with several leptons and jets in the final state.
Previous studies\cite{AMET,SLEP} have shown that the clean, acollinear
$e^+e^-+\eslt$ and
$\mu^+\mu^-+\eslt$ channels offer the best hopes for the discovery of sleptons
at the LHC.
Our main purpose here is to delineate the region of
the SUGRA parameter space where these signals might be observable
at the LHC, and to check whether these can be distinguished from
corresponding signals from chargino pair production.

To separate the signal from SM backgrounds, we require\cite{SLEP},
\begin{itemize}
\item {\it i})~exactly two isolated same flavor OS leptons, each with
$|p_T(\ell)|>20$~GeV,
\item {\it ii})~$\eslt>100$~GeV,
\item {\it iii})~a veto on central
jets with $E_T>25$~GeV within $|\eta|<3$, and
\item {\it iv})~$\Delta\phi(\vec{p_T}(\ell\bar{\ell}),\vec{\eslt})>160^0$.
\end{itemize}

After cuts {\it i}) -- {\it iv}), the dominant SM backgrounds to the SUSY
signal
come from $t\bar{t}$ (2.2~fb) and
$W^+W^-$ (2.9~fb) yielding a ``5$\sigma$ observability level"
of 3.6~fb for a year of LHC operation at the design luminosity.
The slepton cross section is, however, rather small and a higher reach
is obtained with somewhat stiffer cuts to further reduce
the background at modest cost to the signal. Hence,
\begin{itemize}
\item {\it v}) for detection of heavy sleptons, we also require
$|p_T(\ell)|>p_T^c$ and $\Delta\phi (\ell\bar{\ell})<\Delta\phi_c$,
where $p_T^c$ and $\Delta\phi_c$
can be adjusted appropriately. In our analysis, we fix $p_T^c=40$~GeV
and $\Delta\phi_c=90^0$.
\end{itemize}
Including cut {\it v}), we find no events pass the cuts from our
simulation of the $WW$ sample (a one event level corresponds to
$\sigma=0.0015\,\fb$ in our simulation) while from
$t\bar{t}$ events, we find a background cross section of $0.07\pm
0.006\,\fb$.
We thus expect $< 1$~background event per LHC year, with our
``hard'' cuts {\it i}) -- {\it v}).

The region of the $m_0$ -- $\mhf$ plane where slepton production should
yield observable signals after these hard cuts is shown in Fig.~\ref{SLH},
again for the same four cases as in previous figures. Since the
SM background level is very small, we show contours of constant
cross sections corresponding to the 3 -- 5 event level per LHC year
(triangles),
5 -- 10 event level (hollow squares), 10 -- 20 event level (squares with
crosses)
and $>20$ event level (filled squares). The crosses denote
the sampled points for which the cross section is smaller than 3~fb.
We also show contours where $m_{\tell_R}=$~70, 100, 150, 200 and 250~GeV.
If we take the five event level to give the optimistic reach, we see
that the reach of the LHC extends to $m_{\tell_R}\sim 250$~GeV, corresponding
to $m_{\tell_L}$ and $m_{\tnu}$ to just over
$300$~GeV for larger values of $\mhf$. For a SM background expectation
of 0.7 events, the Poisson probability of a fluctuation to the five (ten)
event level is  $8\times 10^{-4}$ ($4\times 10^{-9}$),
so  that a conservative estimate of the reach after a year of LHC operation
is somewhere between 5 and 10 events.

More disturbing is the existence
of the ``hole'' where the cross section falls below the five event
level for small values of $m_0$ and $\mhf$ in cases {\it a}) and {\it c}).
Notice that unless the energy of LEP2 is upgraded so as to
ensure the detectability of sleptons as heavy as 100~GeV,
$\tell_R$ (and, of course, also $\tnu$ and $\tell_L$) may evade detection
at both LEP2 as well as at the LHC.
To understand why  the hole is much larger for the $\mu < 0$ cases,
we have examined the differences in sparticle
properties for ($m_0,\mhf$) = (40~GeV, 140~GeV) in cases {\it a}) and {\it b}).
For the negative $\mu$ case~{\it a}),
$\tw_1$ and $\tz_{1,2}$ are somewhat heavier than in case~{\it b}) so that
the mass difference between $\tell_R$
and $\tz_1$ is rather small (14~GeV, in our example).
As a result, the efficiency for particularly $\tell_R\bar{\tell_R}$
events to pass the {\it hard} $p_T(\ell)>p_T^c$
and $\eslt$ cuts is reduced, leading to a drop in the
cross section. For case {\it b}) $m_{\tell_R}-m_{\tz_1}=30$~GeV so
that the daughter leptons are considerably harder. In addition, $\tell_L$
predominantly decays to $\tw_1$ and $\tz_2$, and further, the leptonic
branching fraction for $\tw_1$ is enhanced to 22\%, while the neutralino
decays via $\tz_2\to\tell_R\ell$, so that hard leptons
can come via several chains. This example also underscores the importance
of incorporating the various cascade decays into the slepton analysis.

We have just seen that because of the hard cuts that we have used in
Fig.~\ref{SLH}, there are small regions of parameter space where
sleptons with masses $\sim 80$ -- 120~GeV
may evade detection both at LEP2 and at the LHC.
Because of the importance of this issue, we have re-done our analysis
using just cuts {\it i}) -- {\it iv}) for which the SM background
cross section is 5.1~fb. In Fig.~\ref{SLS} we show the regions
of the $m_0$-$\mhf$ plane where the significance of the signal
$\sigma=N_{\rm signal}/\sqrt{N_{\rm bkgd}}$ is
$3\sigma$ (triangles), $5\sigma$ (hollow squares), $10\sigma$ (squares
with crosses) and $20\sigma$ (filled squares)
for the same four cases as in Fig.~\ref{SLH}. Indeed we see that
with the soft cuts, the slepton signal always exceeds the $5\sigma$
limit in the ``hole'' regions of Fig.~\ref{SLH}, and further, that
there is no window of masses where sleptons will escape detection
both at LEP2 and at the LHC. The maximal reach at the LHC is, of course,
obtained using the {\it hard} cuts.

In order to check whether dilepton events from slepton pair production
might be confused with corresponding events from chargino production,
we have checked the origin of the events which satisfy our cuts for
several cases:
\begin{itemize}
\item hard cuts, with ($m_0,\mhf$)= (210~GeV, 160~GeV), case~{\it a}) for
which the slepton masses are 225 -- 250~GeV and $m_{\tw_1}=155$~GeV;
\item hard cuts with ($m_0,\mhf$)= (60~GeV, 160~GeV), case~{\it b}) for
which the slepton masses are 92 -- 135~GeV and $m_{\tw_1}=109~$~GeV;
\item hard cuts with ($m_0,\mhf$)= (40~GeV, 140~GeV), case~{\it b})
for which the slepton masses are 75 -- 115~GeV and $m_{\tw_1}=86~$~GeV.

\end{itemize}

In all these cases,
although we had generated all slepton (including sneutrino) as well as
$\tw_1\overline{\tw}_1$ events using ISAJET,
we found that only slepton events in our
sample of 40 -- 60 events that pass our cuts
{\it i.e.} there were no events from direct chargino pair production
in the sample. We did find events from cascade decays of sneutrino (produced
in pairs or along with a charged slepton) as
well as $\ttau$s.
To check whether chargino production contaminates the slepton
sample with the soft cuts {\it i}) -- {\it iv}) in Fig.~\ref{SLS},
we have checked the sources for case~{\it a}) with
($m_0,\mhf$) = (40~GeV, 140~GeV), for which the cross section is
$<0.3\,\fb$ after hard cuts, but where the signal exceeds $5\sigma$
with soft cuts.
We find that out of a total of about forty events
that pass the cuts in our simulation, just six come from direct
chargino production, with the charginos decaying via $\ell\tnu_{\ell}$,
($\ell=e,\mu$).
We thus conclude that a conclusive observation of
a dilepton signal with the hard cuts
will be unlikely to be confused with chargino pair production.
There may, however, be some small
chargino contamination of the signal with the soft
cuts. In this case, the event sample should be large enough to provide
other handles on chargino-slepton discrimination.
For instance,
if $m_{\te_L}=m_{\tmu_L}$ chargino production should lead to
as many $e^{\pm}\mu^{\mp}$ events as $e^+e^- + \mu^+\mu^-$ events,
whereas we would expect significantly more same flavor events
in the case of slepton pair production.

Finally, for the first two cases with the hard cuts above, as well as for the
the soft cut case we just discussed, we generated {\it all} SUSY subprocesses
and ran them through the ``slepton cuts'' to see whether the ``slepton signal''
is contaminated by squark and gluino production, which occurs with much
larger cross section. This requires a simulation of a very large
number of events since only a very tiny fraction of events pass the
cuts. We examined the twenty events that satisfied the ``slepton cuts''
in each of these three cases: we found just
one event from squark and gluino sources in one of the three
event samples. However, in almost half the events
for the $\mu < 0$, hard cut and the soft cut cases, the leptons
both originated
from $\tz_2$ decays in $\tw_1\tz_2$ or $\tz_2\tz_2$ events (the leptons
from the decay of a single $\tz_2$ satisfy the $\Delta\phi$ cut
more readily than those from $\tw_1\tw_1$ events). In the $\mu > 0$,
hard cuts case simulated, we have $m_{\tell_L} > m_{\tw_1},m_{\tz_2}
>m_{\tell_R}$, so that $\tz_2$ always decays via $\tz_2 \to \ell\tell_R$
into real sleptons: for this case, we found about 80\% of the events
had their origin in $\tw_1$ and $\tz_2$ production. We thus conclude
that while squark and gluino production is unlikely to contaminate
the slepton sample, $\tz_2$ decays from $\tw_1\tz_2$ or $\tz_2\tz_2$
production can significantly contaminate the slepton signal (presumably
$\tw_1\tw_1$ events frequently fail the $\Delta\phi$ cut, which fails
to remove dileptons from $\tz_2$ decays). However,
these processes will themselves lead to characteristic signatures (the
clean trilepton signature discussed above or even 4$\ell$ topologies)
and would be detectable in their own right. We should also add
that since these have not been included in Fig.~\ref{SLH} and Fig.~\ref{SLS},
the actual cross sections may be somewhat larger than shown in these
figures.

Before drawing final conclusions regarding the detectability of
sleptons at the LHC, we stress that we have
assumed a 100\% jet rejection efficiency
for jets in the fiducial region. A real detector will, of course,
have cracks and other dead regions. This is especially important
here because the crucial cut\cite{SLEP}
for the detectability
of sleptons over the background from $t\bar{t}$ production is
the central jet veto. In our previous analysis\cite{SLEP}, we
had shown that with the hard cuts, the $t\bar{t}$ background
increases by about a factor of about five if instead this veto efficiency
is 99\%. Except to point out that it may be possible to
reduce this detector-dependent background significantly by
adjusting $p_T^c$ and $\Delta\phi_c$, we will not discuss this
any further. We thus conclude that if detectors have the
capability to veto central jets with a high efficiency, it
should be possible to probe $\tell_R$ and $\tmu_R$ masses up to
about 250~GeV at the LHC. The slepton signals are, however, very small so that
perhaps 20 -- 30~$\fb^{-1}$ of integrated luminosity may be necessary
to confidently probe their existence.

Finally, we point out that in a recent paper\cite{BBDM},
the cosmological relic density from
neutralinos produced in the early universe was evaluated for the same
SUGRA model. In these calculations, it was found that a relic density of
$\Omega h^2 \sim 0.15-0.4$, which is favored by cosmological models
with a critical density and a 2:1 mixture of cold/hot dark matter,
would occur mainly if the
slepton mass $m_{\tell_R}\sim 100-250$ GeV. Thus, failure to detect a slepton
at LHC could place rather severe constraints on cosmological scenarios
which ascribe the bulk of cold
dark matter in the universe to stable neutralinos.

\section{Comparison of Results from the Various Channels}

We have used ISAJET to map out the region of parameter space of the
minimal SUGRA model with radiative breaking of
electroweak symmetry where various $n$-jets plus $m$-leptons ($n \geq 2$,
$m=1,2,3$) plus
$\eslt$ signals are observable above SM backgrounds at the LHC. These
signals are dominantly expected to come mainly from
gluinos and squark production followed by cascade decays.
This paper is a continuation of our previous study\cite{BCPT}
where we had focussed on multijet plus $\eslt$ events with
an isolated lepton veto to reduce backgrounds from vector boson and top quark
production. We also examined the reach in the complementary clean
dilepton and trilepton
channels to investigate the detectability of the electroweak production
of sleptons and charginos/neutralinos at the LHC.

Since the parameter space of the model is rather large, it is
impractical to sample all regions of this space. One approach
would be to generate random sets of model parameters
($m_0$, $\mhf$, $\tan\beta$, $A_0$, $\sgn \mu$) and investigate
various signals for the set of models thus obtained. This is the
strategy used in Ref.\cite{MRENNA} where the authors generated about $2K$
parameter sets in their exploration of the SUSY
reach of the Tevatron and its possible upgrade options. While this is
indeed a viable strategy and may indeed have the advantage that it
samples the parameter space ``more uniformly'', it has some shortcomings.
First, one has to choose how to sample each direction; {\it e.g.} should
one randomly generate $m_0$ or $\ln m_0$, since the measure on parameter
space is unknown. This is important because (for each sign of $\mu$)
just $1000^{\frac{1}{4}}=5.6$ points are generated on average along each
of the four directions.
Second, and more importantly in our view, while it
is true that there may well be a fairer sampling of parameter
space with this approach,
it is difficult to relate the results to the underlying parameters of
the theory. For these reasons, we have chosen to perform detailed
scans in the $m_0$ -- $\mhf$ plane (sparticle masses which dominantly
determine the rates and distributions of the various signals are most
sensitive to these parameters) for fixed values of $\tan\beta$ and $A_0$.
We illustrate the results for a small ($\tan\beta=2$) and a medium
($\tan\beta=10$) value of $\tan\beta$. We do not consider larger values
of $\tan\beta$ because the effects of bottom and tau Yukawa interactions,
which could become important, have not yet been completely included in
ISAJET. In most of our analysis, we fix $A_0=0$ (this does not mean that
the weak scale value of the $A$-parameter vanishes) since our signals
are moderately insensitive to this choice (see Fig.~\ref{AZERO})
except very close to the boundaries
of the parameter space region where the correct pattern of electroweak
symmetry breaking is not obtained.

The details of our calculation in the multijet channels may be found
in Sec.~III, while the clean multilepton signals are discussed in
Sec.~IV. Instead of repeating this discussion one more time, we have
chosen to summarize the results for the LHC reach in the various
channels in Fig.~\ref{SUMM} for $A_0=0$ and ({\it
a})~$\tan\beta=2,\mu<0$, ({\it b})~$\tan\beta=2,\mu>0$, ({\it
c})~$\tan\beta=10,\mu<0$, and ({\it d})~$\tan\beta=10,\mu>0$. As
before, the hatched (bricked) regions are excluded by experimental
(theoretical) constraints.  For a signal to be regarded as
observable\cite{REACHFN}, we require that for an integrated luminosity
of $10\,\fb^{-1}$ at the LHC, the event rates and numbers satisfy:
\begin{itemize}
\item a statistical significance $\geq 5\sigma$, where
$\sigma = N_{\rm signal}/\sqrt{N_{\rm bkgd}}$\,;
\item $N_{\rm signal}/N_{\rm bkgd}\geq 0.2$;
\item $N_{\rm signal} \geq 5$.
\end{itemize}
In the region Fig.~\ref{SUMM} below the dashed line (labeled $\eslt$)
the 0 lepton plus $\eslt$ signal should be observable beyond the
$5\sigma$ level for an appropriate choice of the cut variable $E_T^c$
defined in Sec.~III as well as in Paper I\cite{BCPT} from which these
contours have been taken. The various dashed-dotted contours mark the
boundaries of the region where the $1\ell$, same-sign (SS) dilepton,
opposite-sign (OS) dilepton and trilepton ($3\ell$) plus multijet plus
$\eslt$ signals should be observable at the LHC, again for some value
of $E_T^c \leq 200$~GeV (400~GeV in the case of the 1$\ell$ signal) as
obtained from the analysis in Sec.~III.  The regions below the dotted
line (labeled $\tell$) and solid line (labeled $\tw_1\tz_2$) are where
the clean dilepton and trilepton signals are observable as discussed
in Sec.~IV.

Several comments are worth noting:
\begin{itemize}

\item At the LHC, it should be possible to detect gluinos as heavy as
1.5 -- 1.8~TeV ($m_{\tg} \sim 2.3$~TeV if $m_{\tq} \simeq m_{\tg}$),
corresponding to $\mhf \leq 600$ -- 700~GeV, after
just one year of running at its lower design luminosity option of
$10\,\fb^{-1}/{\rm year}$. This is considerably beyond\cite{FN6} the bounds
($\mhf \alt 400$~GeV) obtained
from (admittedly subjective) fine-tuning arguments, and so should
provide some safety margin for the detectability of SUSY
at the LHC, at least within this minimal framework with conserved
$R$-parity. We also remark that we found no holes where
these signals (or the multilepton signals, for that matter)
might escape detection.

\item It is interesting that the maximal reach is obtained in the $1\ell$
channel. This is because there are numerous sources of leptons in SUSY
events so that a lepton veto significantly reduces the signal cross section.
Our analysis using the $E_T^c$ parameter shows that backgrounds from
$W$ boson and $t\bar{t}$ production (which lead to isolated
leptons in the final
states) can be controlled  without vetoing events with
leptons. It should thus be
possible to combine the signals in the $\eslt$ and $1\ell$ channels to
obtain a somewhat larger reach.

\item If squarks and gluinos are lighter than 1~TeV, several other
signals should be observable above SM backgrounds if a signal in the
$\eslt$ or $1\ell$ channels is to be attributed to sparticle production.
Although our conclusion, strictly speaking, has been obtained in the
rather constrained SUGRA framework, including constraints from radiative
electroweak symmetry breaking (these essentially fix $|\mu|$), previous
analyses\cite{EARLY} suggest that this will be true even if constraints
from electroweak symmetry breaking are relaxed\cite{FN7}. We also
note that a portion of the multileptonic signals arise from leptonically
decaying $Z$ bosons. This is the reason why the reach in multilepton
channels is slightly larger in the $\tan\beta=10$ cases ({\it c}) and ({\it d})
in Fig.~\ref{SUMM}. The real $Z$ boson signals are sensitive to the value
of $\mu$\cite{EARLY}, and hence, to the radiative symmetry breaking
constraint.

\item While it appears that only a rather small subset
of the parameter plane can be probed via the clean leptonic channels, the
observation of these signals is important because it leads to direct
detection of $\tw_1$, $\tz_2$ (this sparticle may be hard to detect
even at the NLC) and the sleptons. Moreover, it has been shown \cite{TRILHC}
that it is possible to isolate $\tw_1\tz_2\to 3\ell$ events
from SM backgrounds as well as from other SUSY sources. This allows
for a reliable determination of $m_{\tz_2}-m_{\tz_1}$, and perhaps, other
combinations of chargino and neutralino masses. We stress that the
non-observation of a trilepton signal at the LHC will not lead
to a bound on the chargino or neutralino mass because of parameter space
regions where the leptonic decays of $\tz_2$ are strongly suppressed.
It is, however, interesting to note that even in these regions, the
multi-jet plus $3\ell$ signals are observable, implying that there are
significant other sources of leptonic events (notably, third generation
fermions and sfermions).

\item At the LHC, it should be possible to detect sleptons with masses
up to 250~GeV (300~GeV for $\tell_L$) in the clean OS dilepton channel.
We have also shown that sleptons as light as
80~GeV ought to be detectable at the LHC
using the ``soft cuts'' discussed in Sec.~IV. Thus, there is
no window where sleptons might escape detection, both at LEP2 and at the
LHC. Furthermore, the LHC is sensitive to the most favored range of slepton
masses expected from calculations of the dark matter neutralino relic
density ($m_{\tell_R}\sim 100-250$ GeV)\cite{BBDM}.

\end{itemize}

Aside from the question of the detection of SUSY, it is interesting
to ask whether it is possible to devise tests of the various assumptions
underlying the minimal SUGRA framework that we have adopted for our
analysis. Tests that work well at an electron-positron collider \cite{MUR,JLC}
do not appear to be feasible at the LHC, partly because the initial state
of the colliding partons is not known, and partly because of the messy
interaction environment at the LHC. Alternatively, we may ask
whether it is possible to use the multitude of observables that should
be accessible at the LHC to determine the underlying parameters of the model.
This is clearly a complex task since the directly observable
quantities such as cross sections in various channels depend
on various masses and mixing angles which have to be unraveled
in order to get at the underlying parameters.

In this paper, we have made a first attempt to understand how
it might be possible to use the LHC data to get at $m_0$ and $\mhf$.
We have little to say at present about the determination
of $\tan\beta$, $A_0$ or $\sgn \mu$.

\begin{enumerate}
\item If $\mhf \alt 300$~GeV (so that gluinos
are lighter than about 700 -- 800~GeV)
we had shown in Paper I that it should be possible to measure $m_{\tg}$ to
15-25\% by requiring hemispheric separation of events in the $\eslt$ channel.
Presumably, the same strategy can also be used in the $1\ell$ channel.
The value of $m_{\tg}$ can be directly related to $\mhf$ (aside from the
(usually small) corrections due to differences between the running and pole
gluino masses).

\item If the trilepton signal from $\tw_1\tz_2$ production is observed at
a substantial rate, it would be possible to check whether
the value of $m_{\tz_2}-m_{\tz_1}$ is in agreement with the expectation
from the gluino mass, assuming that $|\mu|$ is large and the unification
condition for gaugino masses is valid. If the gluino and neutralino
masses are not in accord with this expectation, we would probably conclude
that $|\mu|$ is not large which would imply that we are somewhat
close in parameter space the boundary of the bricked region where
the correct pattern of electroweak symmetry breaking is not obtained.
The alternative would be that the gaugino mass unification condition
is invalid.

\item If the various multijet signals are
observed at rates compatible with gluino masses corresponding to $\mhf \leq
200$~GeV, but no clean trilepton signal is seen, we would probably infer
that we are in one of the ``hole'' regions where the leptonic decays
of $\tz_2$ are strongly suppressed. This would imply that $m_0$, and hence,
squarks and sleptons cannot be too heavy (although there would be no
guarantee that sleptons would be light enough to be observable.)

\item If a signal is observed in the OS dilepton channel with the ``slepton
cuts'' of Sec.~IV, we would place ourselves in the bottom left corner
below the dotted line in the $m_0$-$\mhf$ plane. In this case the multijet
topologies from gluino and squark production {\it must} be seen. Otherwise,
the assumptions of universal sfermion and/or gaugino
mass at the ultra-high scale,
which imply $m_{\tq}^2=m_{\tell}^2+(0.7-0.8)m_{\tg}^2$ could
not be valid.

\end{enumerate}

It is possible that gluinos are rather heavy so that neither the $\tw_1\tz_2$
nor the slepton signals are accessible. The determination of parameters is more
difficult in this case. We have shown in Sec.~III that the cross sections
for multijet plus lepton signals will place us on one of the contours
in Fig.~\ref{ONEL},~\ref{OSL},~\ref{SSL} or~\ref{TRIL}.
Because the multilepton contours are roughly horizontal
(except in the $m_0 \leq 400$ -- 500~GeV region, to which we will come back
to),
it should be possible to get a rough idea of $\mhf$
(roughly within $\pm (50\hbox{ -- }100)$~GeV)
and hence, of $m_{\tg}$. It should also be possible to decide whether
$m_0$ is small ($\alt 300-400$~GeV) or rather large, with a degree of
confidence by studying the ratio of $\sigma(0\ell+\jets) /
\sigma(n\ell+\jets)$  for $n=1$ -- 3. For small values of $m_0$ and
somewhat large values of $\mhf$,
the leptonic decays of charginos and neutralinos, and hence the multilepton
signals, are enhanced. For the same reason, the $0\ell+\jets$ signal (because
of the lepton veto) is reduced, as can be seen by the down turn of the
corresponding contour in Fig.~\ref{SUMM}. This could be confirmed by
a measurement of the flavor asymmetry in the OS dilepton sample (see
Fig.~\ref{FLAS}).
If gluinos are heavy, and
$m_0 \geq 500$~GeV, the determination of $m_0$ may be more difficult.
Possible handles are the charge asymmetry in the $1\ell$ and SS event
samples (the asymmetry reduces with $m_0$) or the jet and $B$ multiplicities
in the 0, 1 and 2 lepton multijet samples (the multiplicity is larger
for larger values of $m_0$). Clearly detailed case studies beyond the scope
of this analysis would be required to determine how well these model parameters
can be determined.

We have not found any strategies for the determination of $\tan\beta$,
$A_0$ or $\sgn\mu$.
A qualitative idea of whether $\tan\beta$ is small
(close to unity) or large might be obtained by looking for multijet events
with real $Z$ bosons: these are more abundant for larger values of $\tan\beta$.
The observation of the Higgs boson and a measurement of its mass (perhaps
in the $\gamma\gamma$ channel) may
also provide a handle on this parameter: since $m_{H_l}=0$ at tree level
if $\tan\beta=1$, the lightest
Higgs boson tends to be lighter when $\tan\beta$ is
close to unity. The parameter $A_0$ mainly affects the third generation.
Variations in $A_0$ can alter significantly the dominant gluino decay
channels, so that rates in di- and, especially,
trilepton plus multijet channels can have
significant dependence on this parameter.
The multiplicity of central $B$-hadrons in SUSY
events should also be sensitive to $A_0$.
We have, however, not studied this
aspect of parameter space in enough detail to draw any clear conclusions.
on $A_0$.

It may well be that all the parameters will
ultimately be extracted by a global fit to all the data. The success
of such a fit would certainly be non-trivial since the complete set
of observations would need to be fitted by just four parameters (plus
a sign). If an adequate fit is not possible, the assumptions
underlying the model would need re-examination.

To sum up, if supersymmetry is the new physics that ameliorates
the fine tuning problem of the SM, it appears almost certain that
there will be a multitude of new physics signals at the LHC. Although
our analysis has been performed within the framework of the $R$-parity
conserving minimal SUGRA model, we do not expect the results to
be qualitatively altered due to minor modifications of the model,
as long as $R$-parity is conserved. The maximal reach is obtained
in the single lepton channel and it appears that gluinos as heavy
as 1.5 -- 1.8~TeV (2.3~TeV if squarks are degenerate with gluinos)
ought to be detectable at the LHC with just $10\,\fb^{-1}$ of data.
It should also be possible, in
at least some cases, to identify the sparticle origins of various
signals. We have also made a preliminary exploration to see how one might
attempt to localize
the underlying SUGRA model parameters, given that these SUSY signals are
seen at the LHC. While this may well be easier at $e^+e^-$ colliders (with
sufficient center of mass energy), it is certainly worthwhile
to think about what might be possible in experiments at the LHC, where
construction has already been approved.
We have argued that it might be possible to extract
$\mhf$ and, to some extent, also $m_0$ via a simultaneous study of
several signals. Other parameters appear even more difficult to obtain,
but this study should only be regarded as a first attempt in this
direction.

%
\acknowledgments
One of us (XT) is grateful to the High Energy Physics Group
at Florida State University for their generous hospitality while
this work was being carried out. In addition, CHC thanks the Davis
Institute for High Energy Physics.
This research was supported in part by the U.~S. Department of Energy
under contract number DE-FG05-87ER40319, DE-FG03-91ER40674,
DE-AC02-76CH00016, and DE-FG-03-94ER40833.
%
%

\newpage
%
%

\iftightenlines\else\newpage\fi

\begin{table}
\caption[]{Results of background calculation in fb after cuts using cut
parameter $E_T^c=200$ GeV. We list the hard scattering $p_T$ ($p_T^{HS}$)
ranges over which the background processes were evaluated,
and then the backgrounds from various SM processes. The upper bounds
quoted correspond to the one event level.
We take $m_t =170$ GeV.}

\bigskip

\def\d{\phantom{0}}

\begin{tabular}{lccccc}
$p_T^{HS}$ & $t\bar t$ & $QCD$ & $W$+jets & $Z$+jets & $WW+WZ+ZZ$ \\
\tableline
$1\ell$ & & & & & \\
$50-100$   & $<0.64$ & $<391$  & $<4.8$ & $<0.74$ & $<0.07$ \\
$100-200$  & $<1.0$  & $<26$   & $<1.0$ & $<0.17$ & $<0.02$ \\
$200-400$  & $3.7$   & $<1.5$  & $7.8$  & $0.26$  & $<0.003$ \\
$400-800$  & $7.2$   & $<0.05$ & $5.7$  & $0.33$  & $0.011$ \\
$800-1600$ & $0.42$  & $0.02$  & $0.39$ & $0.02$  & $0.002$ \\
$1600-3200$& $0.001$ & $0.0004$& $0.004$& $0.0$   & $0.0$ \\
OS & & & & & \\
$50-100$   & $<0.64$ & $<391$  & $<4.8$ & $<0.74$ & $<0.07$ \\
$100-200$  & $1.0$   & $<26$   & $<1.0$ & $<0.17$ & $<0.02$ \\
$200-400$  & $2.6$   & $<1.5$  & $0.61$ & $1.0$   & $<0.003$ \\
$400-800$  & $2.1$   & $0.19$  & $0.52$ & $0.63$  & $0.011$ \\
$800-1600$ & $0.06$  & $0.001$ & $0.02$ & $0.01$  & $0.002$ \\
$1600-3200$& $0.0$   & $0.0001$& $0.001$& $0.0$   & $0.0$ \\
SS & & & & & \\
$50-100$ & $<0.04$  & $<391$  & $<0.42$ & $<0.35$ & $<0.07$ \\
$100-200$  & $<0.05$& $<26$   & $<0.08$ & $<0.08$ & $<0.02$ \\
$200-400$  & $<0.02$& $<1.5$  & $0.09$  & $0.009$ & $<0.003$ \\
$400-800$  & $0.02$ & $<0.05$ & $0.11$  & $0.007$ & $<0.0004$ \\
$800-1600$ & $0.001$& $<0.001$& $0.009$ & $0.0002$& $0.0$ \\
$1600-3200$& $0.0$  & $0.0$   & $0.0$   & $0.0$   & $0.0$ \\
$3\ell$ & & & & & \\
$50-100$   & $<0.04$ & $<391$  & $<0.42$ & $<0.35$  & $<0.07$ \\
$100-200$  & $<0.05$ & $<26$   & $<0.08$ & $<0.08$  & $<0.02$ \\
$200-400$  & $<0.02$ & $<1.5$  & $0.02$  & $<0.009$ & $<0.003$ \\
$400-800$  & $0.01$  & $<0.05$ & $0.03$  & $0.01$   & $<0.0002$ \\
$800-1600$ & $0.002$ & $<0.001$& $0.002$ & $0.0004$ & $0.0$ \\
$1600-3200$& $0.0$   & $0.0$   & $0.0$   & $0.0$    & $0.0$ \\
\end{tabular}
\end{table}



\begin{figure}
\caption[]{Contour plots of squark and gluino masses in the
$m_0$ -- $\mhf$ plane of the minimal SUGRA model. Frames are shown for
{\it a}) $\tan\beta =2,\ \mu <0$, {\it b}) $\tan\beta =2,\ \mu >0$,
{\it c}) $\tan\beta =10,\ \mu <0$, and {\it d}) $\tan\beta =10,\ \mu >0$.
We take $m_t=170$ GeV and $A_0=0$. The bricked regions are excluded by
theoretical constraints discussed in Paper~I,
while the shaded regions are excluded by experiment.}
\label{SQGL}
\end{figure}

\begin{figure}
\caption[]{Same as Fig.~1, except we plot contours of lightest chargino
mass and contours of right slepton mass. Also shown by dotted contours
are the kinematic limit for the neutralino spoiler decay modes, above
which the decays
$\tz_2\to\tz_1 H_{\ell}$ or $\tz_2\to \tz_1 Z$ are kinematically allowed.}
\label{WINSL}
\end{figure}

\begin{figure}
\caption[]{SM backgrounds to various SUSY search event topologies in fb,
after cuts, but as a function of the cut parameter $E_T^c$ defined
in the text. We show
frames for {\it a}) $1\ell+\jets$ events, {\it b}) OS dilepton + jets events,
{\it c}) SS dilepton + jets events and {\it d}) $3\ell +\jets$ events.}
\label{BACK}
\end{figure}

\begin{figure}
\caption[]{SUSY signal cross sections for six SUSY cases listed in the text,
and total SM background in fb, after cuts, as a function
of the $E_T^c$ parameter, for the same event topologies as in Fig.~3.}
\label{SIGNAL}
\end{figure}

\begin{figure}
\caption[]{Contours of cross section  (in fb) after cuts described
in the text for
$1\ell +\jets+\eslt$ events. The solid contours have $E_T^c=100$ GeV,
while the dashed contours are for $1,2$ and $4$~fb cross sections with
$E_T^c=400$~GeV, from which the maximum reach is derived.
The frames are for the same SUGRA parameter choices as in Fig.~1.}
\label{ONEL}
\end{figure}

\begin{figure}
\caption[]{Charge asymmetry $A_c$ defined in the text of the isolated
lepton in
$1\ell+\jets+\eslt$ events. We have fixed $A_0=0$ and $\tan\beta =2$.}
\label{L1AS}
\end{figure}

\begin{figure}
\caption[]{The mean jet multiplicity $\langle n_{j} \rangle$ in
$1\ell+\jets+\eslt$ events for the same cases as in Fig.~6.
We have $A_0=0$ and $\tan\beta =2$.}
\label{L1JM}
\end{figure}

\begin{figure}
\caption[]{The mean tagged $b$-hadron multiplicity $\langle n_B \rangle$ in
$1\ell+\jets+\eslt$ events for the same cases as in Fig.~6.
We have fixed $A_0=0$ and $\tan\beta =2$. The tagging
requirements are described in the text.}
\label{NB1L}
\end{figure}

\begin{figure}
\caption[]{Contours of cross section (in fb) after cuts for
OS $\dilepton + \jets + \eslt$ events. The solid contours have
$E_T^c=100$ GeV,
while the dashed contours are for $1,2$ and $4$~fb cross sections with
$E_T^c=200$ GeV, from which the maximum reach is derived.
The frames are for the same SUGRA parameter choices as in Fig.~1.}
\label{OSL}
\end{figure}

\begin{figure}
\caption[]{Flavor asymmetry ($A_F$) defined in the text
for the OS $\dilepton +\jets +\eslt$ event sample for the same
parameters as in Fig.~\ref{OSL}. We use $E_T^c=200$~GeV (denoted
by diamonds) except when $\mhf =100$~GeV for which we use $E_T^c=100$~GeV
(denoted by squares). The hollow (filled) symbols denote 0.2$\leq A_F \leq 0.5$
($A_F \geq 0.5$), while crosses show the points sampled for which $A_F < 0.2$,
which is consistent with zero in our simulation.}
\label{FLAS}
\end{figure}

\begin{figure}
\caption[]{Same as Fig.~9, except for SS $\dilepton + \jets
+\eslt$ events.}
\label{SSL}
\end{figure}

\begin{figure}
\caption[]{Charge asymmetry $A_c$, defined in the text,
of SS dileptons in
$\ell^{\pm}\ell'^{\pm}+$jets$+\eslt$ events.
We have fixed $A_0=0$ and $\tan\beta =2$.}
\label{SSAS}
\end{figure}

\begin{figure}
\caption[]{Same as Fig.~9, except for $3\ell+\jets+\eslt$ events.}
\label{TRIL}
\end{figure}

\begin{figure}
\caption[]{An illustrative example showing the
variation in cross section after cuts versus the SUGRA
parameter $A_0$, for $1\ell$, SS, OS and $3\ell +\jets+\eslt$ events.
Other SUGRA parameters are listed in the figure. We take $E_T^c=100$ GeV.}
\label{AZERO}
\end{figure}

\begin{figure}
\caption[]{Regions of the $m_0\ vs.\ \mhf$ plane where clean
(central-jet vetoed) isolated tri-lepton events are likely to be observable
at the LHC,
assuming $10\,\fb^{-1}$ of integrated luminosity. The frames are the same as in
Fig.~1, except for the $\mhf$ scale limits.
The filled boxed correspond to a $10\sigma$ effect above background,
open boxes to a $5\sigma$ effect, and crosses correspond to sampled points
which were not observable with $10\,\fb^{-1}$. In addition, the kinematic
boundary for various $\tz_2$ two-body decays are shown.}
\label{TRILEP}
\end{figure}

\begin{figure}
\caption[]{Regions of the $m_0\ vs.\ \mhf$ plane where clean
(central-jet vetoed) isolated dilepton events (usually from slepton pair
production) are likely to be visible using {\it hard slepton cuts} described
in the text,
assuming $10\,\fb^{-1}$ of integrated luminosity. The frames are the same as in
Fig.~1, except for the scale limits.
The various symbols correspond to the cross section levels after cuts
listed on the figure. The estimated SM background level is 0.07~fb.
The solid contours correspond to $m_{\tell_R}=70,$ 100, 150, 200 and 250 GeV,
increasing from the lower left.}
\label{SLH}
\end{figure}

\begin{figure}
\caption[]{Regions of the $m_0\ vs.\ \mhf$ plane where clean
(central-jet vetoed) isolated dilepton events (usually from slepton pair
production) are likely to be visible {\it using soft slepton cuts},
assuming $10\,\fb^{-1}$ of integrated luminosity. The frames are the same as in
Fig.~1, except for the scale limits.
The various symbols correspond to the cross sections after cuts
at the $<3\sigma$, $(3\hbox{ -- }5)\sigma$, $(5\hbox{ -- }10)\sigma$,
$(10\hbox{ -- }20)\sigma$ and $>20\sigma$ levels.
The solid contours correspond to $m_{\tell_R}=70,$ 100, 150, 200 and 250 GeV,
increasing from the lower left.}
\label{SLS}
\end{figure}

\begin{figure}
\caption[]{A summary of the LHC reach (assuming $10\,\fb^{-1}$ of integrated
luminosity)
in the $m_0\ vs.\ \mhf$ plane for the four cases of Fig.~1
via the various multilepton channels
discussed in this paper.
The dashed-dotted curves show the
maximal LHC reach (obtained for some choice of $E_T^c$)
for the $1\ell$, SS, OS and $3\ell +\jets+\eslt$ signals.
Also shown is the reach
via the complementary clean dilepton (marked $\tell$)
and clean trilepton (marked $\tw_1\tz_2$) channels. The boundary of
the parameter plane that can be probed via
multijet$+\eslt$ events (with no isolated leptons, denoted by $\eslt$) as
obtained in
Ref. \cite{BCPT} shown for comparison as the dashed curve.}
\label{SUMM}
\end{figure}


\begin{references}
\bibitem{FN1} Our use of the terms MSSM and SUGRA is as defined
in Ref.\cite{BCPT}.
\bibitem{FN2} While these studies all assume
that $R$-parity is conserved, it seems unlikely that there will
be no observable signal (especially at Linear Colliders, but probably even at
the LHC\cite{RPV}) even if this is not the case.
\bibitem{FN3} While several of these signals have been computed
before, we are not aware of a computation of {\it all} the signals
within the SUGRA framework. Such a computation is necessary before
we can begin systematic tests of the model.
\bibitem{FN5} The cross section falls rapidly with the hard scattering
$p_T$, so that if the $p_T$ range is not broken up into smaller
intervals, the higher values of $p_T^{HS}$ will not be sufficiently
sampled. Since $E_T^c$ is typically 200~GeV, events with relatively
small values of $p_T^{HS}$ will not pass the cuts making the computation
very inefficient. Breaking up the $p_T^{HS}$ interval into bins will
obviously improve the efficacy of the calculation.
\bibitem{REACHFN} Since our computation of the signals and, especially,
backgrounds (which require tails of distributions to be reliably
simulated) are
uncertain by a factor $\sim$2-3 because of inherent uncertainties associated
with our calculation,
it could be argued that the reach might be somewhat smaller than that shown
in Fig.~\ref{SUMM}. However, the reader can then make a personal assessment
of the reach from the signal cross section contours in Sec.~III.
\end{references}
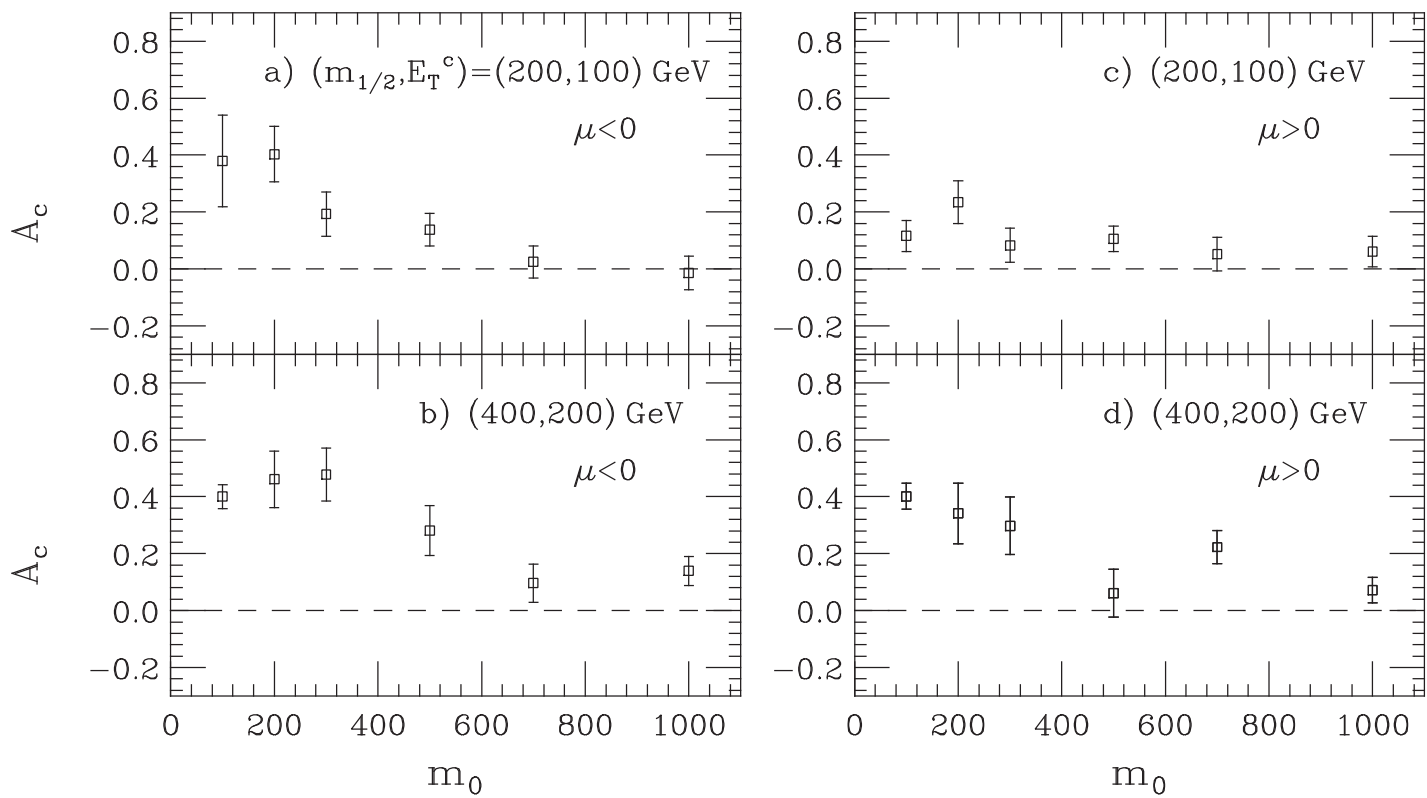
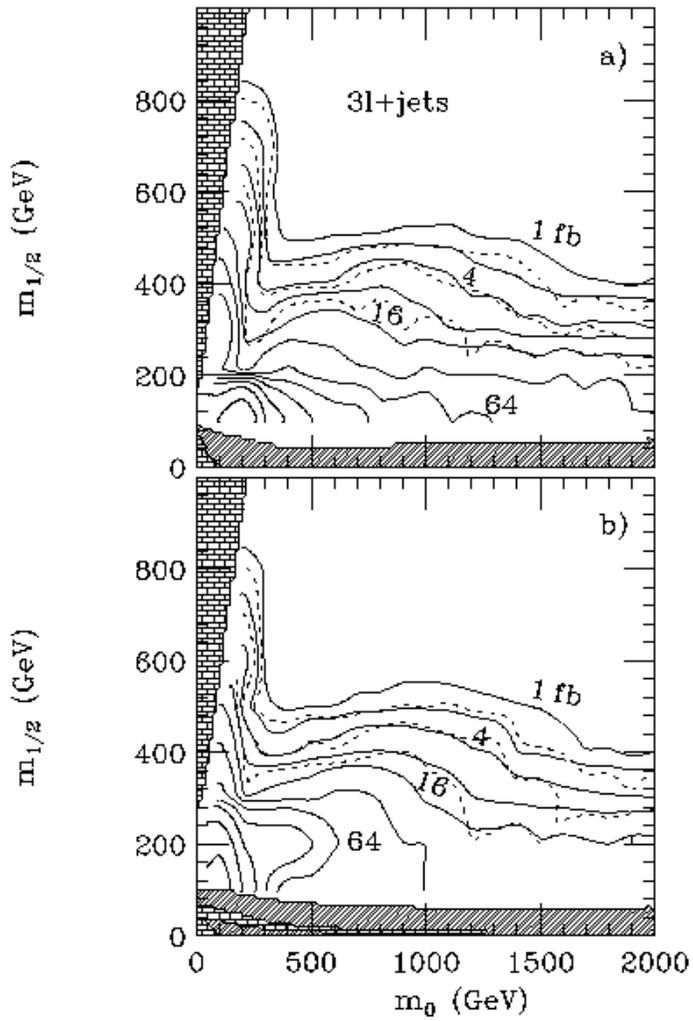
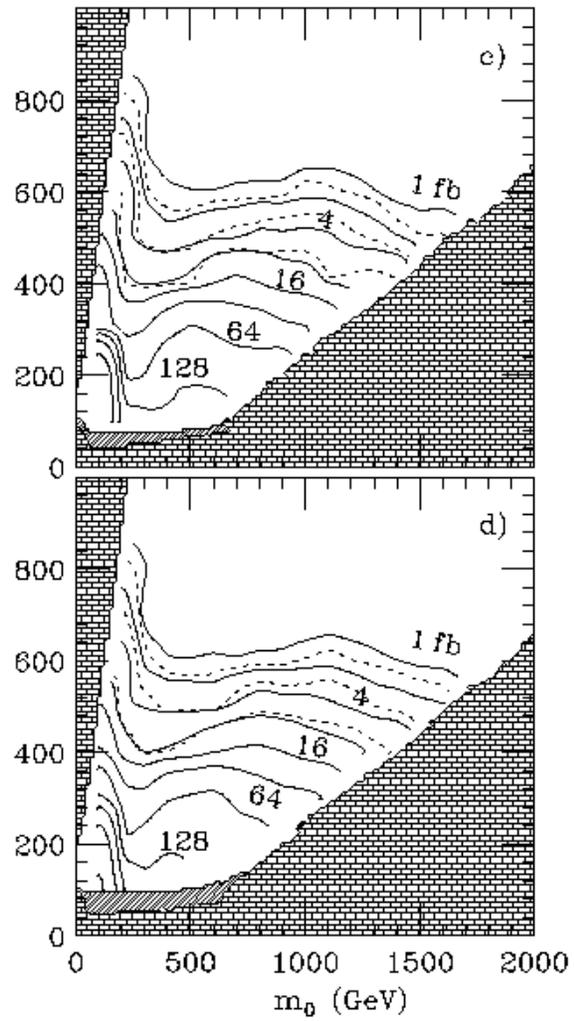
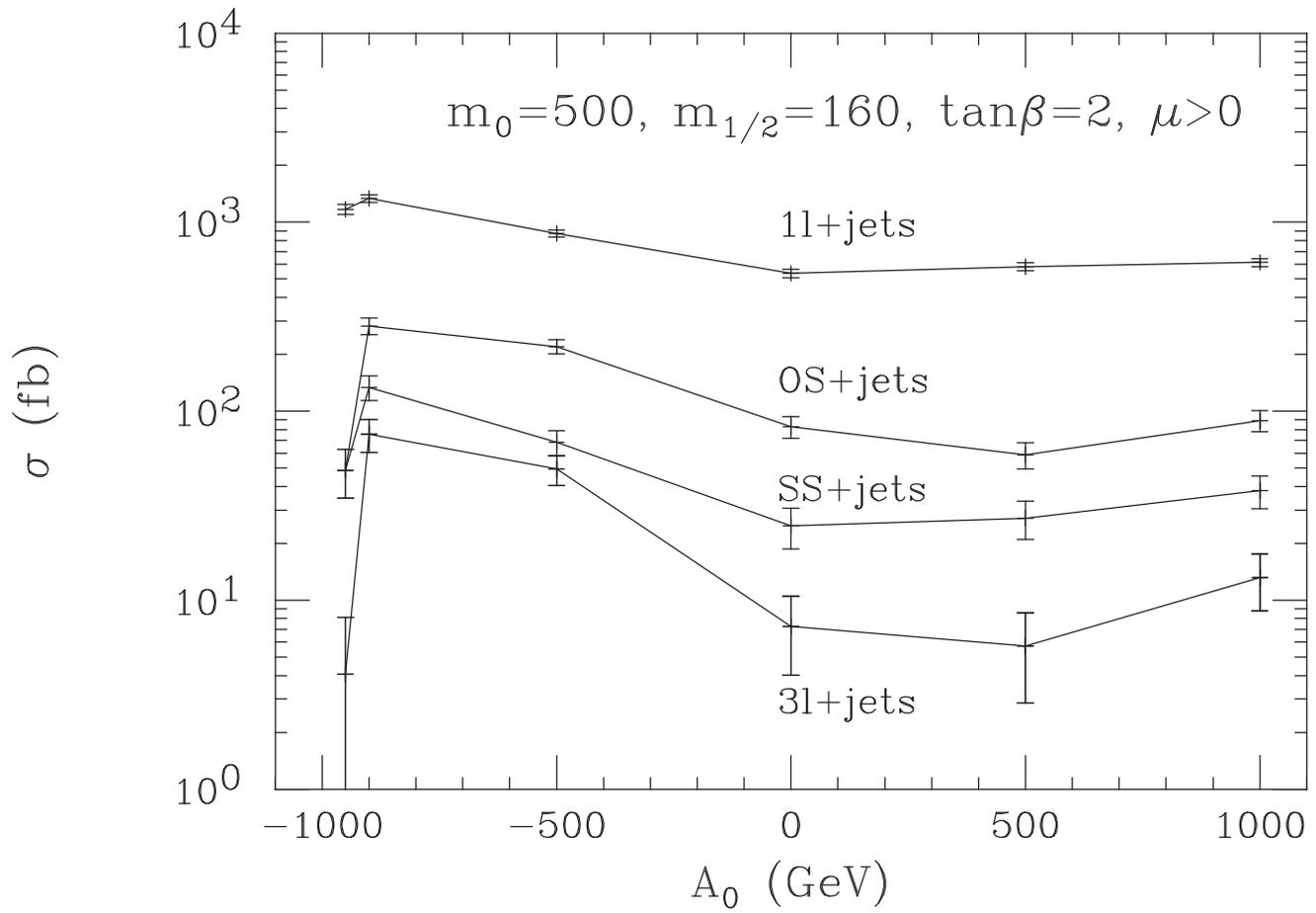
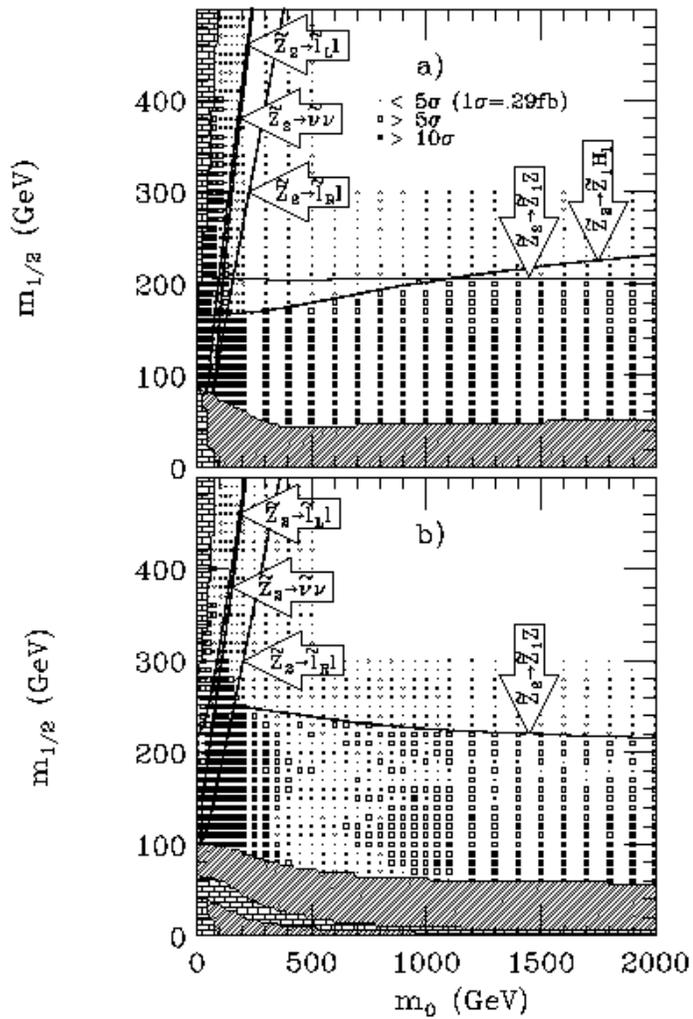
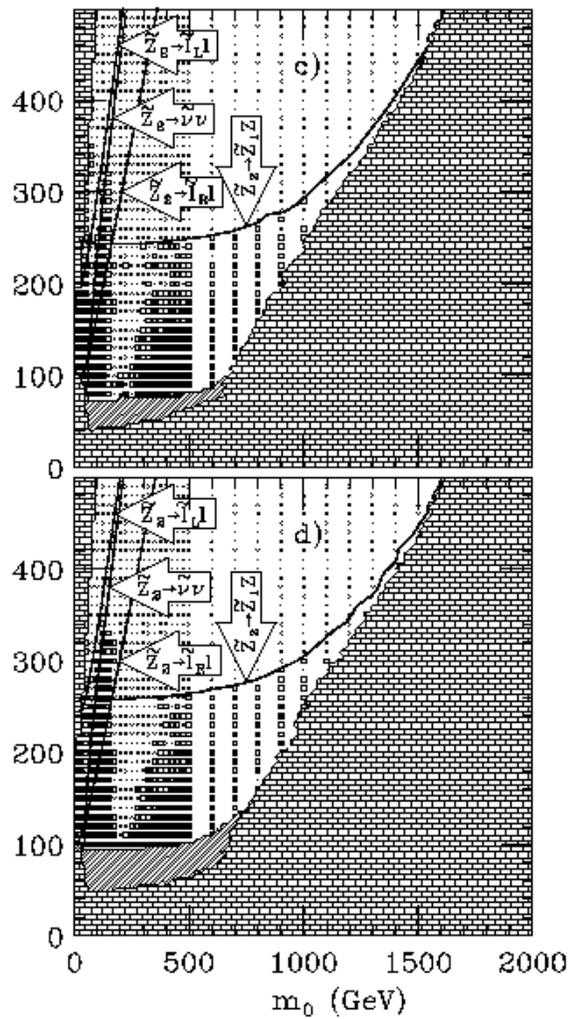
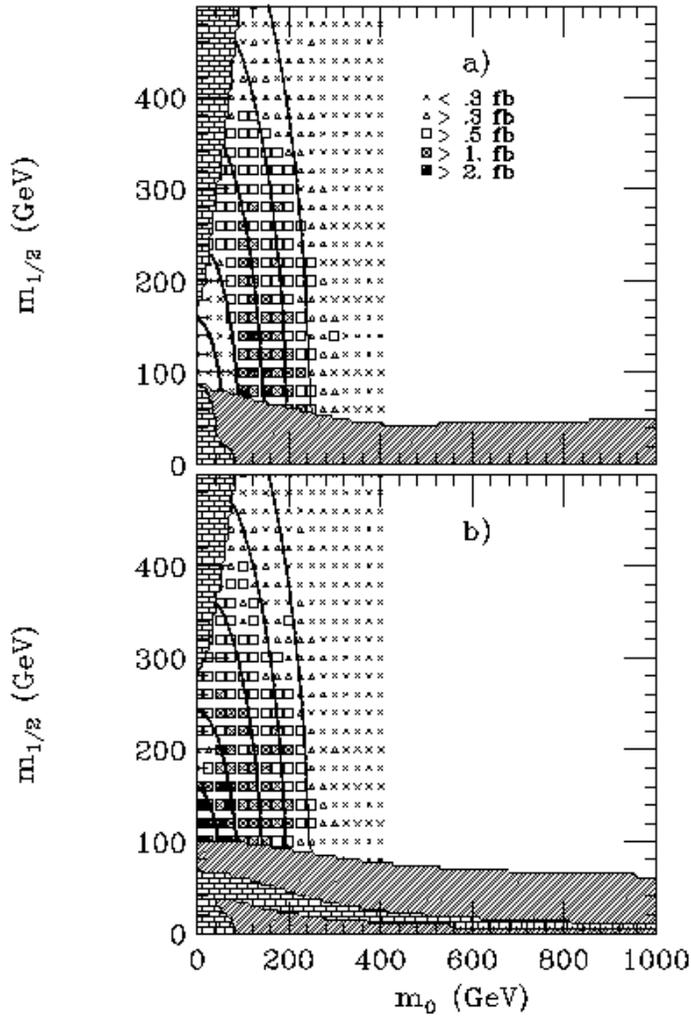
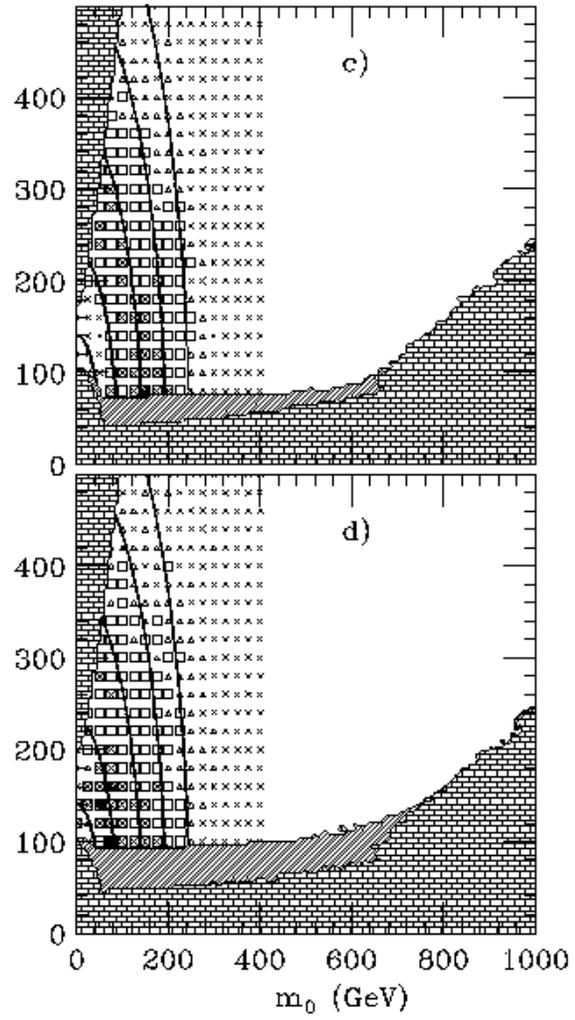
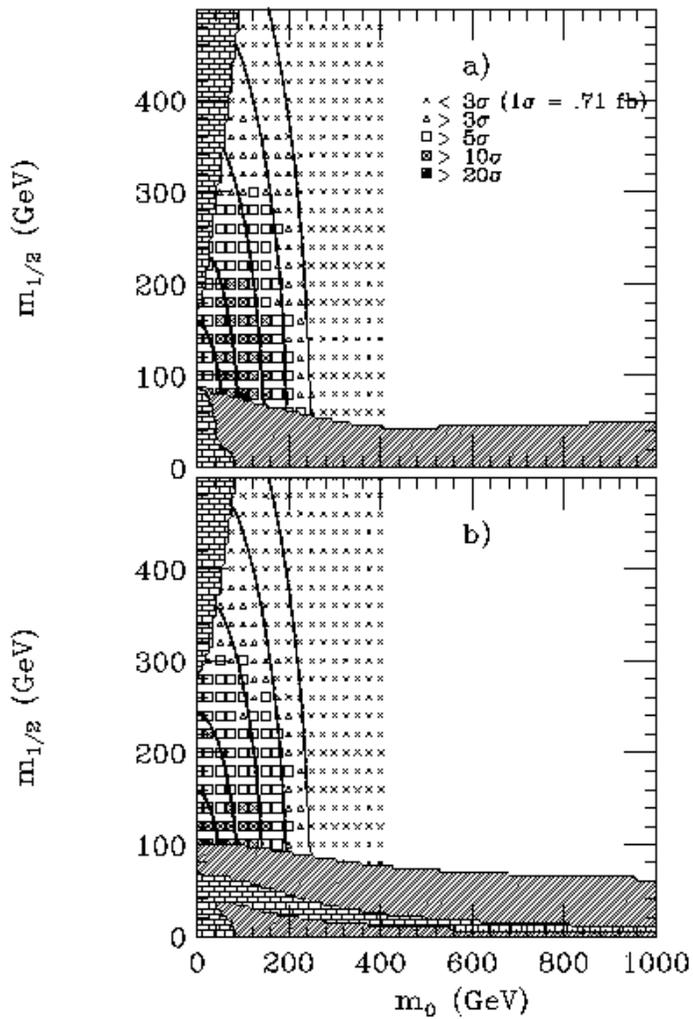
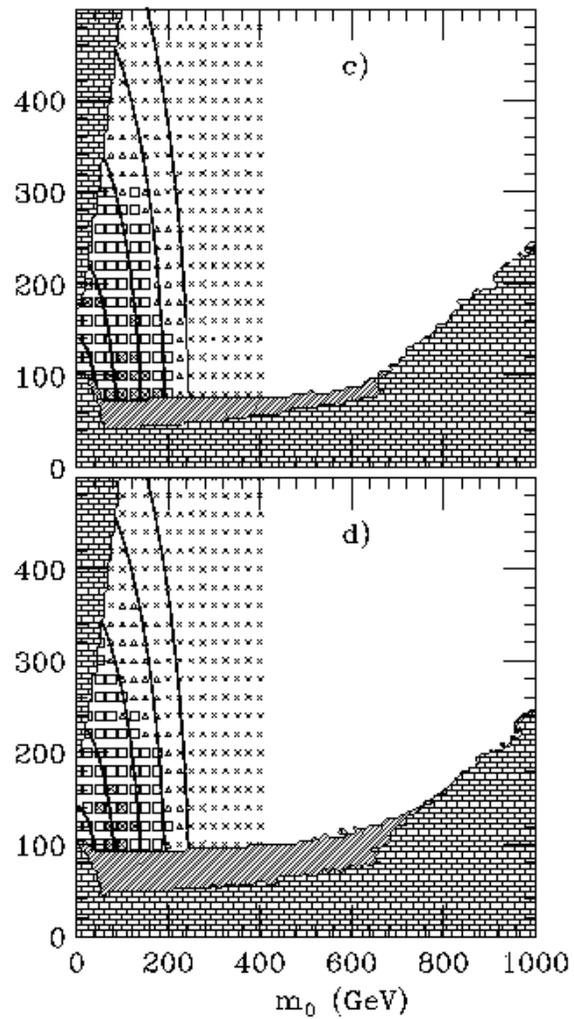
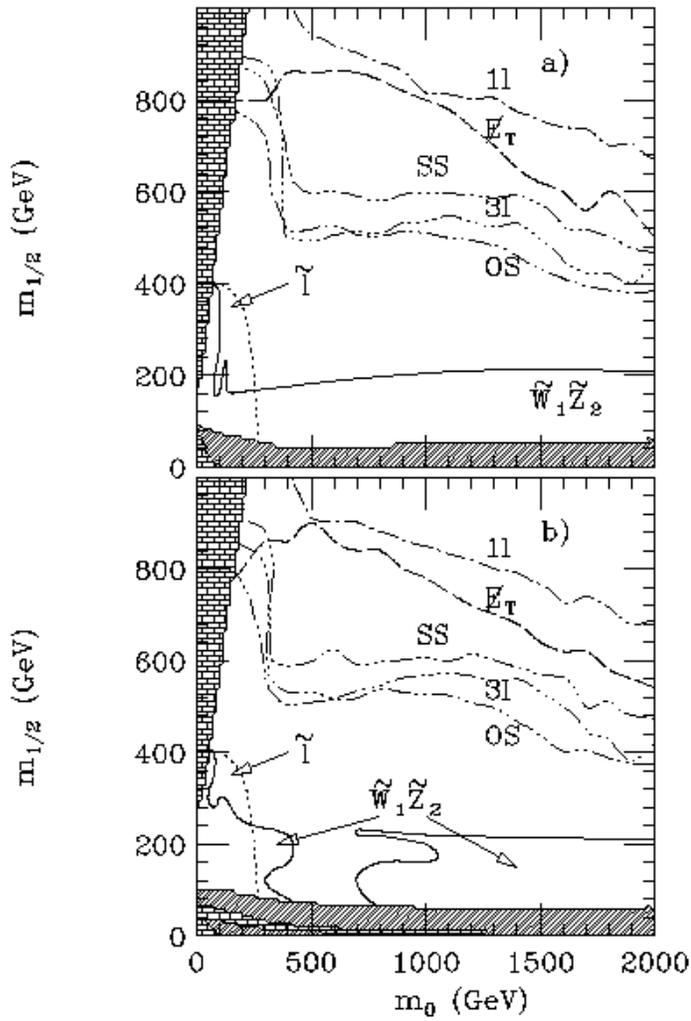
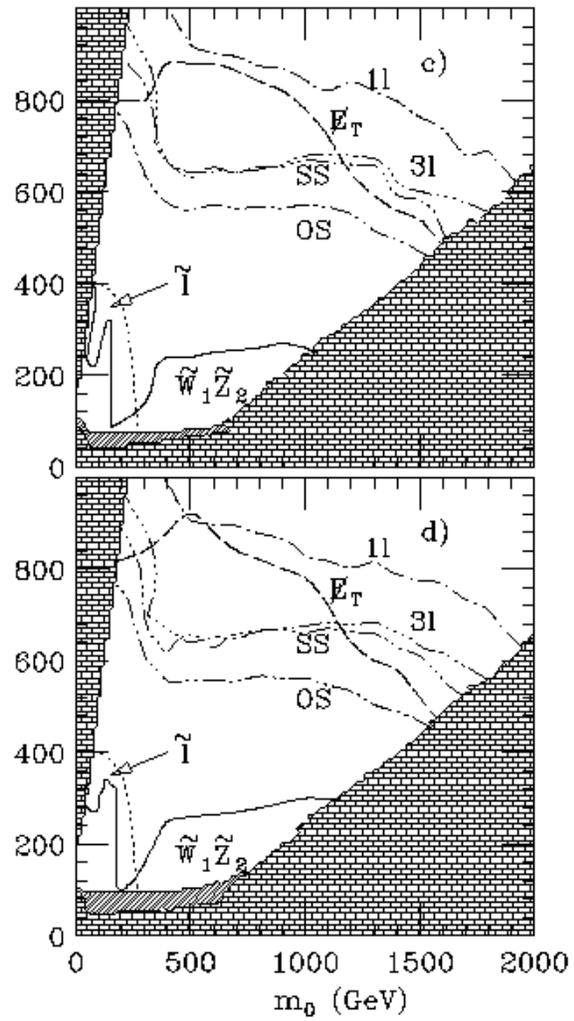
\end{document}